\documentclass[12pt]{article}

\usepackage{graphicx}
\usepackage{amsmath, amssymb}
\usepackage{bm}
\usepackage{cite}
\usepackage{subfigure}
\usepackage[usenames]{color}
  
\newcommand{\LO}{\xi}
\newcommand{\NLO}{\xi^2}

\begin{document}

\begin{titlepage}

\def\thefootnote{\fnsymbol{footnote}}

\begin{center}

\hfill TU-1013\\
\hfill KANAZAWA-16-01\\
\hfill \today

\vspace{0.5cm}

{\Large\bf \boldmath $WW$ scattering in a radiative electroweak
  symmetry breaking scenario }\vspace*{3mm} \\

\vspace{1cm}
{\large Kazuhiro Endo}$^{\it (a)}$, 
{\large Koji Ishiwata}$^{\it (b)}$,
{\large Yukinari Sumino}$^{\it (a)}$

\vspace{1cm}

{\it $^{(a)}${Department of Physics, Tohoku University, Sendai
    980-8578, Japan}}

\vspace{0.2cm}

{\it $^{(b)}${Institute for Theoretical Physics, Kanazawa University,
    Kanazawa 920-1192, Japan}}

\vspace{1cm}

\abstract{ A classically scale invariant (CSI) extension of the
  standard model (SM) induces radiative electroweak symmetry breaking
  and predicts anomalously large Higgs self-interactions.  Hence,
  $W_LW_L$ scattering processes can be a good probe of the symmetry
  breaking mechanism.  We develop a theoretical framework for
  perturbative computation and calculate $WW$ scattering amplitudes in
  a CSI model. It is shown that $W_LW_L$ scattering amplitudes satisfy
  the equivalence theorem, and that a large deviation of $W_LW_L$
  differential cross sections from the SM predictions is predicted
  depending on the c.m.\ energy and scattering angle.  The results are
  more accurate than those based on the effective-potential
  approach. A prescription to implement predictions of the CSI model
  to Monte Carlo event generators is also presented.

}

\end{center}
\end{titlepage}

\renewcommand{\theequation}{\thesection.\arabic{equation}}
\renewcommand{\thepage}{\arabic{page}}
\setcounter{page}{1}
\renewcommand{\thefootnote}{\#\arabic{footnote}}
\setcounter{footnote}{0}

\section{Introduction}
\setcounter{equation}{0}

One of the main targets of the experiments in the second run of the
Large Hadron Collider (LHC) is the weak boson scattering processes in
the TeV energy region, in the hope of finding new physics signals
hidden in the electroweak symmetry breaking sector.  In other words
experimental reach extends to investigating properties of an
``off-shell Higgs boson" through these processes, while thus far our
main focus has been in the investigation of properties of the
``on-shell Higgs boson," where no significant deviations from the
standard model (SM) predictions have been detected.

On the theoretical side there are models with various non-standard
electroweak symmetry breaking mechanisms.  Among them a class of
models with classical scale invariance with extended Higgs sector~
\cite{Espinosa:2007qk,Foot:2007as,AlexanderNunneley:2010nw,Dermisek:2013pta,Antipin:2013exa,Tamarit:2014dua,Endo:2015ifa}
are particularly simple and interesting, in which electroweak symmetry
breakdown is realized via the Coleman-Weinberg
mechanism~\cite{Coleman:1973jx, Gildener:1976ih} at the electroweak
scale.  Due to non-analyticity of the effective potential at the
origin, the vacuum structure of these models is qualitatively
different from that of the SM.  As discussed in
ref.~\cite{Endo:2015ifa}, this is an interesting possibility given the
current status of measurements of the Higgs couplings at the LHC
experiments.  Non-analyticity of the Higgs potential generally leads
to a unique feature different from what one expects from an effective
field theory picture (which assumes expandability of the potential
about the origin).  As a consequence, large deviations of the Higgs
self-couplings from the SM values are predicted, while the Higgs
couplings with other SM particles are barely changed.  In these models
the Higgs cubic coupling is predicted to be larger than the SM values
by a factor 1.6--1.8 and the Higgs quartic coupling by a factor
2.8--4.5~\cite{Dermisek:2013pta,Endo:2015ifa}, which has recently been
confirmed in ref.~\cite{Hashino:2015nxa}.  These models are
perturbatively renormalizable and characterized by a large portal
coupling of the Higgs boson to a non-SM sector.  The size of the
portal coupling is still within the range where perturbative analysis
is valid around the electroweak scale.  Nevertheless, the existence of 
the Landau pole in the region of several TeV to a few tens of TeV 
necessitates an 
UV completion of the models at an energy scale not very far from the
electroweak scale.  A possible scenario of UV completion has also been
proposed in ref.~\cite{Dermisek:2013pta}.

Furthermore, non-SM particles in these models can be part of dark
matter.  In a minimal model detectability of such particles in
experiments of direct detection of dark matter has been
studied~\cite{Endo:2015nba}.  It shows that the model has a parameter
region consistent with the current experimental bounds, which can be
tested in future experiments.

On the other hand, anomalously large self-interactions of the Higgs
boson in this class of models may be detectable in $W$ boson
scattering processes at the LHC experiments.  A rationale is the
equivalence theorem
\cite{Cornwall:1974km,Vayonakis:1976vz,Chanowitz:1985hj}, which states
that scattering cross sections of the longitudinal $W$ bosons
$W_LW_L\to W_LW_L$ approach those of the Nambu-Goldstone (NG) bosons
$GG\to GG$ at high energies.  Since the Higgs boson and NG bosons
compose an $SU(2)_L$ doublet, self-interactions of NG bosons are also
enhanced.

As a first step of an analysis in this direction, in this paper we
take up a minimal model analyzed in ref.~\cite{Endo:2015ifa} and
compute $W$ boson scattering cross sections.  One of our motivations
is to investigate this model as a calculable example of models with a
non-analytic singularity at the origin of the Higgs effective
potential.  We set up a theoretical framework to compute $W$
scattering cross sections at the leading order (LO) of perturbative
expansion.  Due to radiative symmetry breaking, there are non-trivial
theoretical aspects, {\it e.g.}, certain loop corrections need to be
computed in addition to tree-level contributions.  For the computation
a specific order counting needs to be employed as pointed out in
ref.~\cite{Endo:2015ifa}.  In contrast to the effective potential
approach of ref.~\cite{Endo:2015ifa}, we compute by expanding field
components around the vacuum expectation values (VEVs).  In this way
we can compute reliably Feynman amplitudes with non-zero external
momenta.  The explicit calculation of the Feynman amplitudes makes it
possible to discuss details of the kinematics of $W$ boson
scatterings.  As examples, we compute on-shell $WW\to WW$ scattering
amplitudes and cross sections in two channels.  We also check
consistency with the equivalence theorem.

At this stage our computation is somewhat academic since on-shell $WW$
scattering cross sections are difficult to measure realistically. Our
ultimate goal is to perform a feasibility study for testing the model
at the LHC experiments.  For this purpose we need to be able to
implement model predictions to Monte Carlo event generators.  It is
not trivial since the order counting in the Feynman rules is different
from the usual ones and certain loop corrections need to be
incorporated.  In this paper we set a basis for this procedure and
clarify how to implement the model predictions. Besides we compare the
results with the SM computation referring to the past
works~\cite{Duncan:1985vj,Barger:1990py,Denner:1997kq}.

The paper is organized as follows.  In Sec.\,2 we set up necessary
theoretical basis.  Then we compute amplitudes and cross sections for
$WW\to WW$ scattering in Sec.\,3.  Sec.\,4 presents conclusions and
discussion.  Details of the argument and formulas are collected in
Appendices.

\section{Setup}
\label{sec:setup}
\setcounter{equation}{0}

First we present the Lagrangian of the model which we analyze.  The
Higgs potential of the SM is also given, to be used for comparison in
our later discussion (Sec.\,\ref{sec:Lagrangian}).  Then we explain
the order counting used in the perturbative analysis of the model
(Sec.\,\ref{sec:LO&NLO}).  Finally we derive basic relations between
the parameters of the Lagrangian and physical observables, which are
needed for $W$ boson scattering amplitudes
(Sec.\,\ref{sec:Higgssector}).  Consequently the model parameters
needed for our analysis are fixed.

\subsection{Lagrangian}
\label{sec:Lagrangian}

We consider a model, which has an extended Higgs sector with classical
scale invariance (CSI).  Throughout the paper we adopt the Landau
gauge and dimensional regularization with $D=4-2\epsilon$ space-time
dimensions.

The bare Lagrangian of the CSI model is given by
\begin{align}
{\cal L}^{\rm CSI}
=&\,
 	\left[
 		{\cal L}^\text{SM}
 	\right]_{\mu_\text{H}\rightarrow 0}
 	+\frac{1}{2}
(\partial_{\mu}{\vec{S}_B})^{2}
	-{\lambda^{(B)}_\text{HS}}\,
		(H_B^{\dagger}H_B)(\vec{S}_B\cdot\vec{S}_B)
	-\frac{\lambda_{S}^{(B)}}{4}(\vec{S}_B\cdot\vec{S}_B)^2\,.
\label{eq:our_model}
\end{align}
A new real scalar field $\vec{S}=(S_1, \cdots, S_N)^T$ is introduced,
which is a SM singlet and belongs to the $N$ representation of a global
$O(N)$ symmetry.  The above Lagrangian is invariant under the SM gauge
symmetry and the $O(N)$ symmetry.  The singlet field couples to the
Higgs field $H=(H^+, H^0)^T$.  Subscripts or superscripts ``$B$" in
eq.\,(\ref{eq:our_model}) show that the corresponding fields or
couplings are the bare quantities.  The part of the Lagrangian
relevant for the analysis in this paper stems from the Higgs
interaction terms given by
\begin{align}
{\cal L}^{\rm CSI}_\text{H--int}
&=
-\mu^{2\epsilon}(\lambda_{\rm H}+\delta\lambda_{\rm H})(H^\dagger H)^2
-\mu^{2\epsilon}(\lambda_{\rm HS}+\delta\lambda_{\rm HS})
H^\dagger H\,S_iS_i \,.
\label{CSI-HiggsSector}
\end{align}
Here we have re-expressed the interaction terms by renormalized
quantities and counterterms: $H$ and $S_i$ denote the renormalized
fields; $\lambda_{\rm H}$ and $\lambda_{\rm HS}$ represent the
renormalized coupling constants; the terms proportional to
$\delta\lambda_{\rm H}$ and $\delta\lambda_{\rm HS}$ represent the
counterterms; $\mu$ denotes the renormalization scale.

As shown in ref.~\cite{Endo:2015ifa}, the Higgs field acquires a
non-zero VEV via the Coleman-Weinberg mechanism, whereas the singlet
field does not.  We expand the Higgs field about the VEV as $H=(G^+,
(v\mu^{-\epsilon}+h+iG^0)/\sqrt{2})^T$ and set $S_i=s_i$, where $h$,
$G^0$ and $G^+$ represent the physical Higgs, neutral- and charged-NG
bosons, respectively; $v$ denotes the Higgs VEV.  Substituting them into
eq.\,(\ref{CSI-HiggsSector}), one may readily obtain the Feynman rules
for the CSI model.  The tree-level masses of the NG bosons, Higgs
boson and singlet scalar bosons read from the Feynman rules are given
by
\begin{align}
&m_{G^+,{\rm tree}}^2 = m_{G^0,{\rm tree}}^2 =\lambda_{\rm H} v^2 \, , 
\label{eq:mNGtree}  \\
&m_{h,{\rm tree}}^2 = 3\lambda_{\rm H} v^2 \, , 
\label{eq:mhtree}  \\
&m_{s,{\rm tree}}^2 = \lambda_{\rm HS} v^2 \, .
\label{eq:mstree}  
\end{align}
As already noted, certain one-loop corrections can contribute at the
same order as tree-level contributions.  We will see that singlet loop
should be taken into account for determination of the masses of the
Higgs and NG bosons since they contribute at the same order as
$\lambda_{\rm H}$.  Consequently the NG bosons become massless as they
should.  In contrast, the tree-level mass of the singlet scalar bosons
given above corresponds to the physical mass $m_s$ at the leading
order.  These will be shown below, which are also consistent with the
analysis of ref.~\cite{Endo:2015ifa}.

For comparison, the Higgs interaction terms in the SM are given by
\begin{align}
{\cal L}^{\rm SM}_\text{H--int} &=
(\mu_{\rm H}^2+\delta\mu_{\rm H}^2) H^\dagger H
- \mu^{2\epsilon}(\lambda_{\rm H}^{\rm SM}+\delta\lambda_{\rm H}^{\rm SM})
(H^\dagger H)^2 \,.
\end{align}
Note that at tree level
$\mu_H^2=\lambda_{\rm H}^{\rm SM}v^2$, and the 
tree-level Higgs mass is given by
\begin{eqnarray}
(m_{h,\text{tree}}^{\rm SM})^2=2\lambda_{\rm H}^{\rm SM} v^2 \, .
\label{eq:mhSMtree}
\end{eqnarray}
The roles of the Higgs quartic couplings turn out to be quite
different between the CSI model and the SM, hence we distinguish them
as $\lambda_{\rm H}$ and $\lambda_{\rm H}^{\rm SM}$ throughout the
paper.\footnote{ This is not the case for other couplings such as the
  top-quark Yukawa coupling $y_t$ or $SU(2)_L$ gauge coupling $g_2$,
  at least in the LO analysis given in this paper.  }

\subsection{Order counting of parameters}
\label{sec:LO&NLO}

To start our discussion, an important point is that the relation
\begin{eqnarray}
  \lambda_{\rm H} \sim \frac{N\lambda_{\rm HS}^2}{(4\pi)^2} \ll 1
 \label{eq:EWSMcondition} 
\end{eqnarray}
needs to be satisfied for the electroweak symmetry breaking to be
realized via the Coleman-Weinberg mechanism in the perturbative
regime, since tree-level and one-loop effects should balance
\cite{Endo:2015ifa}. Therefore, it is necessary to assign specific
order counting to the parameters of the CSI model within legitimate
perturbation theory.  We clarify the order counting in this model. At
the same time we assign similar specific order counting to the SM so
that we can make clear comparison between the two models. We introduce
an auxiliary expansion parameter $\xi$ and rescale the parameters of
the models as follows:
\begin{align}
  \lambda_{\rm HS}
  \rightarrow
  \LO\, \lambda_{\rm HS} \, ,
~~~
  \lambda_{\rm H}
  \rightarrow
  \NLO\, \lambda_{\rm H} \, ,
~~~
  \lambda_{\rm H}^{\rm SM}
  \rightarrow
  \NLO\, \lambda_{\rm H}^{\rm SM} \, ,
~~~
  \mu_{\rm H}
  \rightarrow
  \LO\,  \mu_{\rm H} ,
~~~
  y_t
  \rightarrow
  \LO^{1/2}\,   y_t ,
\label{eq:coupling-orders}
\end{align}
where $y_t$ denotes the top-quark Yukawa coupling. 
  $\lambda_{\rm HS}\to \xi\, \lambda_{\rm HS} $ is our starting point.
  $\lambda_{\rm H}\to\xi^2\,\lambda_{\rm H}$ and $y_t \to
  \xi^{1/2}\,y_t$ follow from the fact that $\lambda_{\rm H}\sim
  \lambda_{\rm HS}^2/(4\pi)^2\sim y_t^4/(4\pi)^2$ is required for the
  radiative electroweak symmetry breaking in this model, as in
  eq.\,\eqref{eq:EWSMcondition}. For later discussion we have also
  added $\lambda_{\rm H}^{\rm SM}\to\xi^2\,\lambda_{\rm H}^{\rm SM}$
  and $\mu_H\to\xi\mu_H$
  to compare the CSI model with the SM. 
Then we expand each physical observable in series
expansion in $\xi$, and in the end we set $\xi=1$.  Hence, if an
observable is given as $A(\xi)= \xi^{n}(a_0 + a_1\xi + a_2 \xi^2 +
\dots )$, we define the LO term of $A$ as $a_0$, the next-to-leading
order (NLO) term of $A$ as $a_1$, etc.  One may confirm that in this
way the effective expansion parameter becomes
\begin{align}
\xi \sim \frac{\lambda_{\rm HS}}{(4\pi)^2}
\sim \frac{y_t^2}{(4\pi)^2}  \sim
\left[\frac{\lambda_{\rm H}}{(4\pi)^2}\right]^{1/2}
\sim
\left[\frac{\lambda_{\rm H}^\text{SM}}{(4\pi)^2}\right]^{1/2},
\label{eq:ord-xi}
\end{align}
including the loop factor $1/(4\pi)^2$. (See App.~\ref{app:expparam}
for details.)  In particular, since $\lambda_{\rm HS}\lesssim 5$,
$|\lambda_{\rm H}|\lesssim 0.1$, $\lambda_{\rm H}^\text{SM}\approx
0.13$, $y_t\approx 1$, the effective expansion parameter is
sufficiently small to ensure validity of perturbative
expansion~\cite{Endo:2015ifa}.\footnote{ One finds that
  $y_t^2/(4\pi)^2$ is considerably smaller than the other effective
  expansion parameters in eq.\,(\ref{eq:ord-xi}). Nevertheless, we
  treat it as ${\cal O}(\xi)$ since in the relevant cases top-quark
  loops give leading radiative contributions in the SM.  }  In this
first analysis, we compute all the physical quantities at the LO of
the series expansion in $\xi$.

For demonstration, we explicitly write the auxiliary parameter $\xi$
in the following subsection.  It is often useful to note the orders of
the mass parameters in the computation.  We list the orders in $\xi$
of the relevant parameters in Tab.~\ref{table:OC}, where $m_X$ denotes
the physical (on-shell) mass of particle $X$.  The listed orders
follow from the assignment eq.\,(\ref{eq:coupling-orders}) and the
tree-level masses eqs.\,(\ref{eq:mhtree}), (\ref{eq:mstree}),
(\ref{eq:mhSMtree}), provided that loop corrections do not change the
orders of the tree-level masses.  (Indeed this condition holds except
for the masses of the NG bosons.)  We explain computation of the
physical masses in the next subsection.

\begin{table}[t]
\begin{center}
CSI\hspace*{53mm}SM\vspace*{3mm}\\
\begin{tabular}{|c|c|c|} \hline
  \multicolumn{1}{|l|}{Order} &
  \multicolumn{2}{|c|}{Parameters} \\  \hline 
$\xi^1$   & $\lambda_{\rm HS}$, $y_t^2$  & $m_s^2$, $m_t^2$ \\ \hline
$\xi^2$   & $\lambda_{\rm HS}^2$,
$y_t^4$, $\lambda_{\rm H}$ &$m_h^2$ \\
\hline 
\end{tabular} \ \
\begin{tabular}{|c|c|c|} \hline
  \multicolumn{1}{|l|}{Order} &
  \multicolumn{2}{|c|}{Parameters} \\  \hline
$\xi^1$    &  $y_t^2$ & $m_t^2$ \\ \hline
$\xi^2$    & $y_t^4$, $\lambda_{\rm H}^{\rm SM}$, $\mu_{\rm H}^2$
& ${(m_{h,\text{tree}}^{\rm SM})}^2$, $m_h^2$\\
\hline 
\end{tabular}
\caption{\small Orders of the parameters in the CSI model (left) and
  in the SM (right).  }
\label{table:OC}
\end{center}
\end{table}

\subsection{Physical parameters of the Higgs sector}
\label{sec:Higgssector}

The crucial difference between the CSI model and the SM resides in the
Higgs sector.  The Higgs sector of each model determines two
dimensionful parameters, the Higgs VEV and the (on-shell) Higgs mass.
They can be identified as physical parameters\footnote{ Within our
  current approximation (LO in $\xi$ expansion), the Higgs VEV is
  directly related to the Fermi constant by $v=(\sqrt{2}G_F)^{-1/2}$.
} and are determined by the parameters of the bare Lagrangian.  The
relations can be obtained by calculating the Higgs tadpole diagrams
and Higgs self-energy diagrams.

\begin{figure}[t]
\begin{center}
 \includegraphics[scale=0.8]{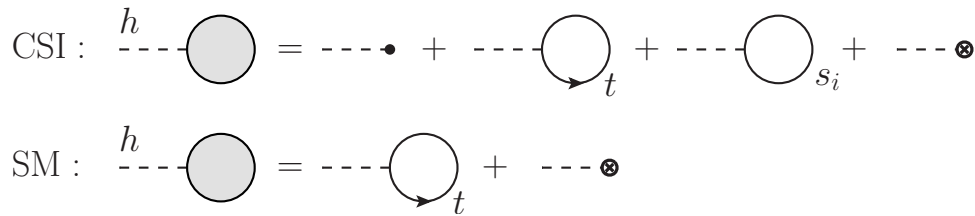}
\end{center}
\vspace*{-3mm}
\caption{\small Higgs tadpole diagrams in the CSI model and the SM.  A
  blob in the CSI model represents the tree-level vertex, while a
  vertex with cross represents the counterterms.  (The counterterms in
  the CSI model and the SM are different. See text.)}
\label{fig:HiggsTadpole}
\end{figure}

In the CSI model, there is a tree-level Higgs tadpole diagram, which
contributes $-\xi^2\lambda_{\rm H} \mu^{-\epsilon}v^3h$, and the
singlet and top-quark one-loop diagrams contribute at the same order.
To cancel the UV divergence, the counterterm is also needed.  In the
SM, on the other hand, the tree-level tadpole contributions cancel
(since we set $\mu_H^2=\lambda_{\rm H}^{\rm SM}v^2$) at ${\cal
  O}(\LO^2)$ and only the counterterms and loop diagrams
remain.\footnote{ We require that the relation $\mu_H^2=\lambda_{\rm
    H}^{\rm SM}v^2$ is unchanged after inclusion of the top-loop
  effect.  In this way we choose a renormalization scheme for the SM
  (at the LO in perturbative expansion in $\xi$), which is suited for
  comparison with the CSI model.  } The corresponding diagrams are
shown in Fig.\,\ref{fig:HiggsTadpole}.  Thus, the conditions for
vanishing tadpole contributions read, respectively, as
\begin{align}
  &{\rm CSI}:~~\NLO\lambda_{\rm H} + \delta \lambda_{\rm H} =
  \NLO\frac{N\lambda_{\rm HS}^2}{(4\pi)^2}\tilde{A}_0(\xi m_s^2)
  - \NLO\frac{N_cy_t^4}{(4\pi)^2}\tilde{A}_0(\xi m_t^2) \, ,
  \label{eq:tadpole^CSI} \\
  &{\rm SM:}~~-\delta \mu_{\rm H}^2/v^2 + \delta \lambda_{\rm H}^{\rm SM} =
  - \NLO\frac{N_cy_t^4}{(4\pi)^2}\tilde{A}_0(\xi m_t^2)\, ,
  \label{eq:tadpole^SM} 
\end{align}
at ${\cal O}(\LO^2)$.\footnote{ We count $\log \xi$ on the right-hand
  side as ${\cal O}(\xi^0)$.  } Here, the top-quark mass is given by
$m_t=y_tv/\sqrt{2}$, the number of colors $N_c=3$, and
$\tilde{A}_0(m^2)$ denotes the loop function defined in
App.~\ref{app:loopfunc}.  Eq.\,\eqref{eq:tadpole^CSI} coincides with
eq.\,(3.11) of ref.~\cite{Endo:2015ifa} in the case that $\mu=v$ and
the counterterm is defined in the $\overline{\rm MS}$ scheme.

\begin{figure}[t]
\begin{center}
  \includegraphics[scale=0.8]{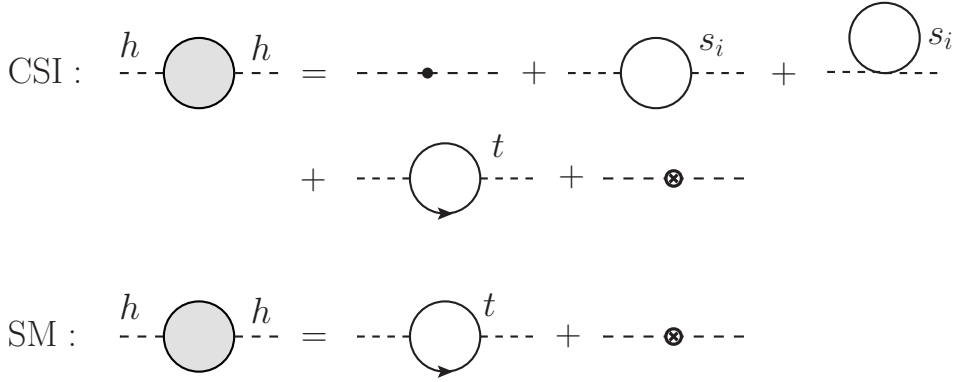}
\end{center}
\caption{\small Higgs self-energy diagrams in the CSI model and the
  SM.  Similarly to Fig.\,\ref{fig:HiggsTadpole}, a blob and a vertex
  with a cross represent the tree-level vertex and counterterm,
  respectively. }
\label{fig:HiggsSelfenergy}
\end{figure}

We can compute the Higgs self-energy in a similar manner.
The corresponding diagrams are shown in
Fig.\,\ref{fig:HiggsSelfenergy}, and
the results are given by
\begin{align}
  &{\rm CSI}:~~\nonumber\\
&
-\Sigma_h(p^2) =
  - 3(\NLO\lambda_{\rm H} +\delta \lambda_{\rm H})v^2+\delta Z_h p^2
  +  \NLO\frac{N\lambda^2_{\rm HS}v^2}{(4\pi)^2}
  \left[2B_0(p^2;\xi m_s^2)+\tilde{A}_0(\xi m_s^2)\right]
  \nonumber \\ &  ~~~~~~~~~~~~~~~~
  - \NLO\frac{2N_cy_t^2}{(4\pi)^2}\left[
    \frac{p^2}{\xi} 
B_1(p^2;\xi m_t^2)+2m_t^2B_0(p^2;\xi m_t^2)+m_t^2\tilde{A}_0(\xi m_t^2)
    \right] + {\cal O}(\LO^3)\, ,
  \label{eq:Sigma_h^CSI} \\
  &{\rm SM:}~~\nonumber\\
&
-\Sigma_h^{\rm SM}(p^2) =
  \delta \mu_{\rm H}^2 -3\delta\lambda_{\rm H}^{\rm SM} v^2+\delta Z_h^{\rm SM} p^2
  \nonumber \\ & ~~~~~~~~~~~~~~~~~
   - \NLO\frac{2N_cy_t^2}{(4\pi)^2}\left[
    \frac{p^2}{\xi} 
 B_1(p^2;\xi m_t^2)+2m_t^2B_0(p^2;\xi m_t^2)+m_t^2\tilde{A}_0(\xi m_t^2)
    \right] + {\cal O}(\LO^3)\, .
   \label{eq:Sigma_h^SM}
\end{align}
Here, $B_i(p^2;m^2)\equiv B_i(p^2;m^2,m^2)$ denote the loop functions
defined in App.~\ref{app:loopfunc}.  We have included the counterterms
for the wave function renormalization $\delta Z_hp^2$, $\delta
Z_h^{\rm SM}p^2$. 
 After the divergence in $B_1(p^2;\xi
  m_t^2)$ is subtracted by $\delta Z_h$, $p^2 \log(p^2)$ remains in
  the self-energy. However, this term is canceled by top-bottom loop
  in $WWh$ coupling. Taking the $WWh$ correction into account, some
  finite pieces give  corrections to the Higgs propagator. However,
  they are $O(\xi^3)$, which is beyond the order of our interest
  in the following discussion. Hence we neglect them hereafter.

Using the self-energy, the on-shell Higgs mass $m_h$ is defined
in each model
as
\begin{align}
  &{\rm CSI}:~~ \NLO m_h^2 = \Sigma_h(\xi^2 m_h^2) \, ,
  \label{eq:mh^CSI} \\
  &{\rm SM}:~~ \NLO m_h^2 = (\xi\,{m_{h,\text{tree}}^{\rm SM}})^2
+\Sigma_h^{\rm SM}(\xi^2 m_h^2) \, .
  \label{eq:mh^SM}
\end{align}
The tree-level SM Higgs mass $m_{h,\text{tree}}^{\rm SM}$
is defined in eq.\,(\ref{eq:mhSMtree}).

For later convenience, we reduce the difference of the Higgs inverse 
propagators of the two models to a simple form.  Combining
eqs.\,\eqref{eq:tadpole^CSI}--\eqref{eq:mh^SM}, we obtain
\begin{align}
 - \Delta \Sigma_h(p^2) \equiv & ~
  \Bigl[ p^2 - \Sigma_h(p^2) \Bigr]
 - \Bigl[ p^2 -
  {(\xi\, {m^{\rm SM}_{h,\text{tree}}})}^2
  - \Sigma_h^{\rm SM}(p^2) \Bigr]
 \nonumber
  \\ =&~
  \NLO \frac{2N\lambda_{\rm HS}^2v^2}{(4\pi)^2}
  \left[B_0(p^2;\xi m_s^2)-B_0(\xi^2 m_h^2;\xi m_s^2)
    \right] \,.
\label{eq:DSigma_h}
\end{align}
Note that the top-loop contributions as well as
contributions of the Higgs quartic couplings have dropped
from this expression.

Let us focus on the CSI model and examine relations between the
parameters of the Lagrangian and physical observables.
Substituting eqs.\,(\ref{eq:Sigma_h^CSI}) and (\ref{eq:tadpole^CSI})
into eq.\,(\ref{eq:mh^CSI}), we obtain a simple expression for the
on-shell Higgs mass as
\begin{align}
\NLO m_h^2=&
-\NLO \frac{2N\lambda^2_{\rm HS}v^2}{(4\pi)^2}
\left[B_0(\xi^2 m_h^2;\xi m_s^2)-\tilde{A}_0(\xi m_s^2)\right]
\nonumber \\ & 
+\NLO \frac{2N_c y_t^4v^2}{(4\pi)^2}
\left[B_0(\xi^2 m_h^2;\xi m_t^2)-\tilde{A}_0(\xi m_t^2)\right] \,
+{\cal O}(\LO^3)
\label{eq:mh}
\\
=&~
\NLO  \frac{2N\lambda^2_{\rm HS}v^2}{(4\pi)^2}
- \NLO  \frac{2N_cy_t^4v^2}{(4\pi)^2}
+{\cal O}(\LO^3)\,,
\label{eq:mhapp}
\end{align}
where in the second equality we used the asymptotic form of
the loop function given in App.~\ref{app:loopfunc},
taking into account $\xi^2 m_h^2\ll \xi m_t^2,\, \xi m_s^2$.

Comparing eq.\,(\ref{eq:mhapp}) with the experimental data, we can
determine $\lambda_{\rm HS}$.  (It is natural to regard this coupling
to be renormalized at scale $\mu\simeq 2m_s$.)  Using the central
values of $m_h=125.03\pm 0.27~{\rm
  GeV}$~\cite{Aad:2014aba,Khachatryan:2014ira}, $v=246.66~{\rm GeV}$
and $m_t=173.34\pm 0.76~{\rm GeV}$~\cite{ATLAS:2014wva}, we
obtain\footnote{ Accuracy of approximating the right-hand side of
  eq.\,(\ref{eq:mh}) by the asymptotic form eq.\,(\ref{eq:mhapp}) is
  within $0.1\,\%$.  }
\begin{eqnarray}
  \lambda_{\rm HS}(\mu\simeq 2m_s)\approx 4.82/{\sqrt{N}} \,.
\end{eqnarray}  
We also find that the top loop
contribution amounts to (only) about $5\%$ in the physical Higgs mass
eq.\,(\ref{eq:mhapp}). 

One can also check that the NG bosons become massless by
similar calculations.

\begin{figure}[t]
\begin{center}
 \includegraphics[scale=0.6]{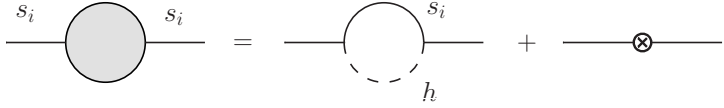}
\caption{\small
Singlet self-energy diagrams contributing to
the lowest-order radiative corrections to the singlet mass $m_s$. 
(Note that $\lambda_S$ is omitted throughout the paper.)}
\label{fig:SingletSelfenergy}
\end{center}
\end{figure}

The singlet mass is given by eq.\,(\ref{eq:mstree}) at tree level. 
The lowest-order radiative correction is given by the singlet-Higgs
one-loop contribution shown in Fig.\,\ref{fig:SingletSelfenergy},
which is ${\cal O}(\NLO)$.
Thus, the physical singlet mass
is given by
\begin{eqnarray}
  \xi m_s^2=\LO \lambda_{\rm HS} v^2 + {\cal O}(\NLO) \,
 \label{eq:ms}
\end{eqnarray}
at the LO.  We summarize the values of the parameters in
Tab.~\ref{table:lambdaHSms}.  They agree well with the previous
results~\cite{Endo:2015ifa}.  We use the values in the table to
compute $WW\to WW$ cross sections in the next section.

\begin{table}[t]
\begin{center}
\begin{tabular}{|c|c|c|c|} \hline
  \multicolumn{1}{|c|}{$N$} &
  \multicolumn{1}{|c|}{1} &
  \multicolumn{1}{|c|}{4} &
  \multicolumn{1}{|c|}{12} 
  \\  \hline 
$\lambda_{\rm HS}(\mu\simeq 2m_s)$ &  4.82 & 2.41 & 1.39 \\ \hline
$m_s$ [GeV]       &  541  & 383  & 291\\
\hline 
\end{tabular} 
\caption{\small Values of the parameters of the CSI model determined
  from the VEV and mass of the Higgs boson.  They are used in the
  computation of $WW\to WW$ cross sections.  }
\label{table:lambdaHSms}
\end{center}
\end{table}

\section{\boldmath $WW$ scattering processes}
\label{sec:amp}
\setcounter{equation}{0}

In this section we investigate scattering processes of the $W$ bosons.
We calculate the amplitudes for the scattering processes $W^+W^-
\rightarrow W^+W^-$ (Sec.\,\ref{sec:W+W-amp}) and $W^+W^+ \rightarrow
W^+W^+$ (Sec.\,\ref{sec:W+W+amp}).  Then we show that in the CSI model
the differential cross sections of the longitudinal $W$ boson
scattering $W_LW_L\to W_LW_L$ deviate from the SM predictions,
especially when the energy scale of the scattering processes is much
higher than the electroweak scale (Sec.\,\ref{sec:WWxsec}).  In this
section we set $\xi=1$ except where we count orders in $\xi$.

\subsection{Amplitude for $W^+W^-\rightarrow W^+W^-$}
\label{sec:W+W-amp}

\begin{figure}[t]
\begin{center}
 \includegraphics[width=16cm]{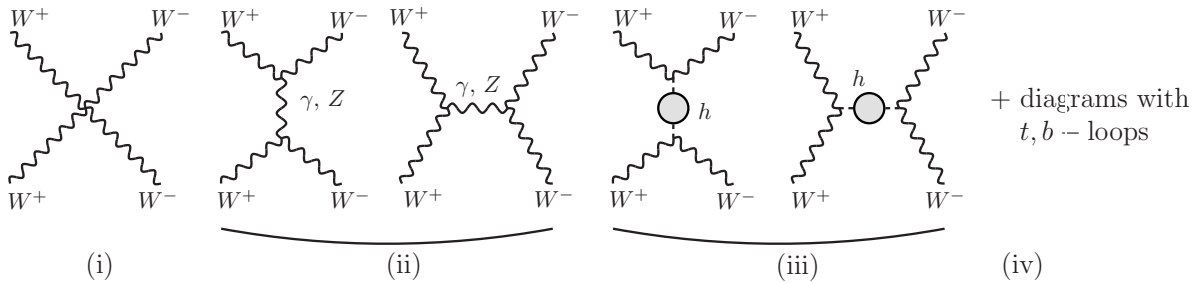}
\end{center}
\vspace*{-7mm}
\caption{\small
Diagrams for $W^+W^- \rightarrow W^+W^-$ scattering. Time flows upwards.}
\label{fig:AmpWpWm}
\end{figure}

First we calculate the scattering amplitude for $W^+W^-\rightarrow
W^+W^-$.  We consider the following four types of diagrams, shown in
Fig.\,\ref{fig:AmpWpWm}: (i) quartic $W$-boson vertex, (ii) $\gamma$-
and $Z$-boson exchange diagrams ($s$- and $t$-channels), (iii)
Higgs-boson exchange diagrams ($s$- and $t$-channels), and (iv)
diagrams including $t$- and $b$-quark loops (up to one loop).  The type
(i) and (ii) diagrams include only tree diagrams.  In the type (iii)
diagrams, we include Higgs self-energy, up to ${\cal O}(\xi^2)$, in
the denominator of the Higgs propagator.  To avoid double-counting, we
eliminate the Higgs self-energy diagram from (iv) .  In the case of
$W_L^+W_L^-\to W_L^+W_L^-$ scattering at high energy, these diagrams
correspond to the LO [${\cal O}(\xi^2)$] amplitude for the NG-boson
scattering $G^+G^-\to G^+G^-$, using the equivalence theorem.

At a first glance, it is not obvious how the above diagrams are
related to the couplings of the Higgs sector, as predicted by the
equivalence theorem.  This is because the couplings of the Higgs
sector do not appear explicitly, except in the Higgs self-energy
diagrams.  As well known, there is a severe gauge cancellation at high
energy among the type (i)--(iii) diagrams.  After gauge cancellation,
the sum of these diagrams behaves proportionally to the Higgs
self-interaction, in accord with the equivalence theorem.  The type
(iv) diagrams are even more subtle.  After gauge cancellation, the
part proportional to $m_t^4$ of these diagrams is expected to give
order $\xi^2 y_t^4/(4\pi)^2$ contributions.\footnote{ By naive
  dimensional analysis, the $W_L$ polarization vectors behave as $\sim
  (E_W/m_W)^4$ ($m_W$ is the $W$ boson mass), and the $m_t^4$ part of
  the rest of the kinematical factors (including loop integrals) as
  $\sim m_t^4/E_W^4$.  Hence, the type (iv) diagrams include the
  behavior $\sim g_2^4 \cdot (E_W/m_W)^4 \cdot m_t^4/E_W^4 \sim
  y_t^4$.  Since positive powers of $E_W$ in these diagrams are
  expected to be canceled due to gauge cancellation, $y_t^4$ part
  would be the dominant part at high energies.  }

The difference of the $W^+W^-\rightarrow W^+W^-$ amplitudes between
the CSI model and the SM originates from the Higgs exchange diagrams
(iii).  We express the amplitude in each model as
\begin{align}
&{\rm CSI}:~~{\cal A}_{W^+W^- \rightarrow W^+W^-}^\text{CSI} =
{\cal A}^{\rm quart}_{W^+W^-}
+{\cal A}^{\gamma/Z}_{W^+W^-}+{\cal A}^h_{W^+W^-} 
+{\cal A}^t_{W^+W^-} \,,
\nonumber \\
&{\rm SM}:~~{\cal A}_{W^+W^- \rightarrow W^+W^-}^{\rm SM} =
{\cal A}^{\rm quart}_{W^+W^-}+{\cal A}^{\gamma/Z}_{W^+W^-}+
  {\cal A}^{h,{\rm SM}}_{W^+W^-} +{\cal A}^{t}_{W^+W^-}\,,
\label{eq:AmpWpWm} 
\end{align}
where ${\cal A}^{\rm quart}_{W^+W^-}$, ${\cal A}^{\gamma/Z}_{W^+W^-}$,
${\cal A}^{h{\rm (,SM)}}_{W^+W^-}$ and ${\cal A}^{t}_{W^+W^-}$
represent the sub-amplitudes corresponding to the diagrams (i)--(iv),
respectively.  ${\cal A}^{\rm quart}_{WW}$, ${\cal
  A}^{\gamma/Z}_{W^+W^-}$ and ${\cal A}^{t}_{W^+W^-}$ are common in
both models, whereas ${\cal A}^{h}_{W^+W^-}$ and ${\cal A}^{h,{\rm
    SM}}_{W^+W^-}$ are different.  As we have seen in the previous
section, the singlet loop gives a LO [${\cal O}(\NLO)$] contribution
to the Higgs self-energy in the CSI model, which is absent in the SM.

Assigning the momenta of the initial- and final-state particles as
$W^+(p_1)W^-(p_2)\rightarrow W^+(k_1)W^-(k_2)$, and setting
$s=(p_1+p_2)^2$ and $t=(p_1-k_1)^2$, we obtain
\begin{align}
&{\rm CSI}:~~{\cal A}^{h}_{W^+W^-}=
  - g_2^2 m_W^2\frac{1}{s-\Sigma_h(s)}\ 
  \epsilon(p_1)\cdot \epsilon^*(p_2)\  \epsilon^* (k_1)\cdot \epsilon(k_2)
  \nonumber \\ &~~~~~~~~~~~~~~~~~~~~~
  -  g_2^2 m_W^2 \frac{1}{t-\Sigma_h(t)}\ 
  \epsilon(p_1)\cdot \epsilon^*(p_2)\  \epsilon^* (k_1)\cdot \epsilon(k_2)
  \,,  
\\
&{\rm SM}:~~{\cal A}^{h,{\rm SM}}_{W^+W^-}=
- g_2^2 m_W^2
\frac{1}{s-({m_{h,{\rm tree}}^{\rm SM}})^2-\Sigma^{\rm SM}_h(s)}\ 
 \epsilon(p_1)\cdot \epsilon^*(p_2)\  \epsilon^* (k_1)\cdot \epsilon(k_2)
 \nonumber \\ &~~~~~~~~~~~~~~~~~~~~~~
 -  g_2^2 m_W^2
 \frac{1}{t-({m_{h,{\rm tree}}^{\rm SM}})^2-\Sigma^{\rm SM}_h(t)}\ 
  \epsilon(p_1)\cdot \epsilon^*(p_2)\  \epsilon^* (k_1)\cdot \epsilon(k_2)
 \,,
\end{align}
where $g_2$ and $m_W$ represent, respectively, the gauge coupling of
$SU(2)_L$ and the $W$ boson mass.  $\epsilon^\mu(p_i)$,
$\epsilon^\mu(k_i)$ represent the polarization vectors of the $W$
bosons characterized by their momenta.  The first and second terms of
each amplitude correspond to the $s$- and $t$-channel Higgs exchange
diagrams, respectively.  Thus, the difference of the two amplitudes
can be attributed to the difference of the Higgs propagators given by
\begin{align}
&\frac{1}{s-\Sigma_h(s)}
- \frac{1}{s-{(m_{h,{\rm tree}}^{\rm SM})}^2-\Sigma^{\rm SM}_h(s)}
\nonumber \\ &~~~~~~~~~~ =
-\frac{1}{s^2}\cdot \frac{2N\lambda_{\rm HS}^2v^2}{(4\pi)^2}\left[
  B_0(s; m_s^2) - B_0(m_h^2;m_s^2)\right]
+{\cal O}(\LO^3)\,,
\label{eq:diff-W+W-}
\end{align}
and to the corresponding difference for the $t$-channel Higgs
propagators.  We have used eq.\,\eqref{eq:DSigma_h}.  Note that
${(m_h^{\rm SM})}^2$, $\Sigma^{\rm SM}_h$, $\Sigma_h$ are all ${\cal
  O}(\xi^2)$ quantities.

The main purpose of our analysis is to clarify the deviation of the
prediction of the CSI model from the SM prediction.  We find that the
deviation can be taken into account by adding the difference
eq.\,(\ref{eq:diff-W+W-}) to each Higgs propagator in the SM amplitude.
Alternatively, one may add $-\Delta \Sigma_h(p^2)$ defined in
eq.\,(\ref{eq:DSigma_h}) to the denominator of the Higgs propagator in
the SM, which is more accurate in kinematical regions close to
on-shell Higgs productions.  This is one of the main results of this
paper.  Noting that eq.\,(\ref{eq:diff-W+W-}) vanishes as $s\to m_h^2$,
we see that indeed an ``off-shell Higgs boson" gives clues to the
electroweak symmetry breaking mechanism, as anticipated in the
Introduction.

Let us check the high energy behavior of the scattering amplitude for
$W_L^+W_L^-\to W_L^+W_L^-$ by comparing to the NG-boson scattering
amplitude.  At high energy $s,|t| \gg m_W^2$, the polarization vectors
of longitudinal $W$ bosons grow.  Consequently, we have
\begin{align}
  &\epsilon_L(p_1)\cdot \epsilon_L^*(p_2)=
  \epsilon^*_L (k_1)\cdot \epsilon_L(k_2)=
  \frac{(\beta^2+1)s}{4m_W^2}
  ~~~~~~~~~~~\to~~\frac{s}{2m_W^2} \,,
  \\
  &\epsilon_L(p_1)\cdot \epsilon^*_L(k_1) =
  \epsilon_L^*(p_2)\cdot \epsilon_L(k_2)=
  \frac{\beta^2(\beta^2-1)s-2t}{4\beta^2m_W^2}
  ~~\to~~\frac{-t}{2m_W^2} \,.
\end{align}
Here, the subscript ``$L$'' stands for the longitudinal mode; $\beta$
is the velocity of the $W$ bosons in the c.m.\ frame, {\it i.e.},
$\beta=\sqrt{1-4m_W^2/s}$. It follows that, at high energy, $s,|t|\gg
m_s^2$, the difference of the scattering amplitudes behaves as
\begin{align}
&{\cal A}_{W^+_LW^-_L \rightarrow W^+_LW^-_L}^\text{CSI}
-{\cal A}_{W^+_LW^-_L \rightarrow W^+_LW^-_L}^{\rm SM}
\nonumber \\ & ~~~
\longrightarrow ~~~
\frac{2N\lambda_{\rm HS}^2}{(4\pi)^2}\left[
  B_0(s;m_s^2) + B_0(t;m_s^2)-2B_0(m_h^2;m_s^2)\right]
\nonumber \\ & ~~~~~~~~~~~~~
\approx
\frac{2N\lambda_{\rm HS}^2}{(4\pi)^2}\left[
  \log \left(\frac{m_s^4}{st}\right)
+4 \right]
+{\cal O}\left(\frac{m_s^2}{s},\frac{m_s^2}{|t|}\right)
\, .
\label{eq:DampWpWm}
\end{align}
This agrees with the difference of 
the $G^+G^-\rightarrow G^+G^-$
amplitudes of the two models,
given in eq.\,\eqref{eq:DAmpGpGm},
which is consistent with the equivalence theorem.

\subsection{Amplitude for $W^+W^+\rightarrow W^+W^+$}
\label{sec:W+W+amp}

\begin{figure}[t]
\begin{center}
 \includegraphics[width=16cm]{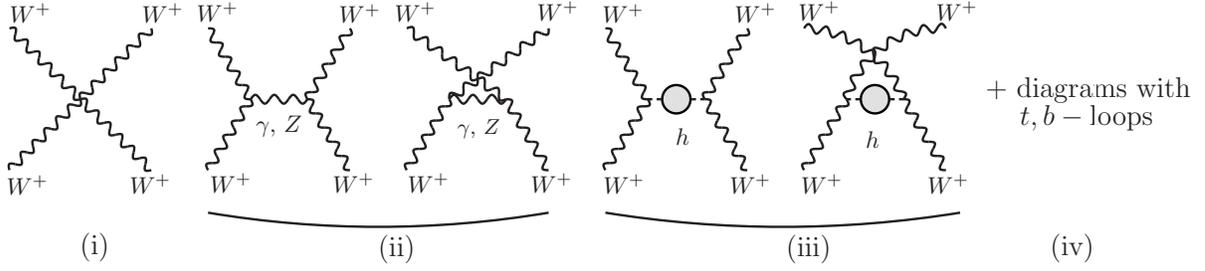}
\end{center}
\vspace*{-7mm}
\caption{\small Diagrams for $W^+W^+ \rightarrow W^+W^+$
  scattering. (Time flows upwards, which is similar to
  Fig.\,\ref{fig:AmpWpWm}.)}
\label{fig:AmpWpWp}
\end{figure}

It is straightforward to compute the $W^+W^+\rightarrow W^+W^+$
scattering process in a similar manner.  The diagrams are shown in
Fig.\,\ref{fig:AmpWpWp}.  The amplitudes for the CSI model and the SM
are given by
\begin{align}
&{\rm CSI}:~~{\cal A}_{W^+W^+ \rightarrow W^+W^+}^\text{CSI} =
{\cal A}^{\rm quart}_{W^+W^+}+{\cal A}^{\gamma/Z}_{W^+W^+}
+{\cal A}^h_{W^+W^+} +{\cal A}^t_{W^+W^+} \,,
\nonumber \\
&{\rm SM}:~~{\cal A}_{W^+W^+ \rightarrow W^+W^+}^{\rm SM} =
{\cal A}^{\rm quart}_{W^+W^+}+{\cal A}^{\gamma/Z}_{W^+W^+}+
  {\cal A}^{h,{\rm SM}}_{W^+W^+}+
  {\cal A}^{t,{\rm SM}}_{W^+W^+} \,,
\end{align}
where the notations are similar to the previous subsection.
The Higgs-exchange diagrams are given by
\begin{align}
&{\rm CSI}:~~{\cal A}^{h}_{W^+W^+}=
  -  g_2^2 m_W^2 \frac{1}{t-\Sigma_h(t)}\ 
  \epsilon(p_1)\cdot \epsilon^*(k_1)\  \epsilon(p_2)\cdot \epsilon^*(k_2)  
  \nonumber \\ &~~~~~~~~~~~~~~~~~~~~~
  -  g_2^2 m_W^2 \frac{1}{u-\Sigma_h(u)}\ 
  \epsilon(p_1)\cdot \epsilon^*(k_2)\  \epsilon(p_2)\cdot \epsilon^*(k_1)
  \,,  
\nonumber \\
&{\rm SM}:~~{\cal A}^{h,{\rm SM}}_{W^+W^+}=
-  g_2^2 m_W^2
\frac{1}{t-({m_{h,{\rm tree}}^{\rm SM}})^2-\Sigma^{\rm SM}_h(t)}\ 
  \epsilon(p_1)\cdot \epsilon^*(k_1)\  \epsilon(p_2)\cdot \epsilon^*(k_2)
\nonumber \\ &~~~~~~~~~~~~~~~~~~~~~~
-  g_2^2 m_W^2
\frac{1}{u-({m_{h,{\rm tree}}^{\rm SM}})^2-\Sigma^{\rm SM}_h(u)}\ 
  \epsilon(p_1)\cdot \epsilon^*(k_2)\  \epsilon(p_2)\cdot \epsilon^*(k_1)
 \,,
\end{align}
with $u=(p_1-k_2)^2$ for $W^+(p_1)W^+(p_2)\rightarrow
W^+(k_1)W^+(k_2)$.
Thus, we can calculate the difference of the two amplitudes
similarly to the previous subsection.

Using
\begin{align}
  \epsilon_L(p_1)\cdot \epsilon^*_L(k_2)=
  \epsilon_L(p_2)\cdot \epsilon^*_L(k_1)=
  \frac{\beta^2(\beta^2+1)s+2t}{4\beta^2m_W^2}
  ~~\to~~ \frac{-u}{2m_W^2} \,,
\end{align}
for $|t|,|u| \gg m_W^2$,
we obtain the high energy behavior of the deviation as
\begin{align}
&{\cal A}_{W^+_LW^+_L \rightarrow W^+_LW^+_L}^\text{CSI}
-{\cal A}_{W^+_LW^+_L \rightarrow W^+_LW^+_L}^{\rm SM}
\nonumber \\ & ~~~
\longrightarrow
\frac{2N\lambda_{\rm HS}^2}{(4\pi)^2}\left[
  B_0(t;m_s^2) + B_0(u;m_s^2)-2B_0(m_h^2;m_s^2)\right]
\nonumber \\ & ~~~~~~~~~
\approx
\frac{2N\lambda_{\rm HS}^2}{(4\pi)^2}\left[
  \log \left(\frac{m_s^4}{tu}\right)+4 \right]
+{\cal O}\left(\frac{m_s^2}{s},\frac{m_s^2}{|t|}\right)
\, .
\label{eq:DampWpWp}  
\end{align}
As expected, this expression agrees with the corresponding amplitude
for $G^+G^+\rightarrow G^+G^+$ given in eq.\,\eqref{eq:DAmpGpGp}.

\subsection{Cross sections for $W_LW_L$ scatterings: Numerical study}
\label{sec:WWxsec}

We perform a numerical study of the $W_LW_L$ scattering cross sections
using the results of the previous subsections.  Up to now, we
considered (at least formally) all the top-loop corrections which
contribute to the LO of $\xi$ expansion at high energy.  In the
following numerical study, however, we include only those part of the
top-loop corrections which are enhanced by logarithms of the energy
$\sim \log s ,\log |t|$, for the following reason.  To the best of our
knowledge, the full one-loop electroweak corrections to the $WW$
scattering processes have been presented only numerically for the
$W^+W^-$ scattering in ref.~\cite{Denner:1997kq} and the analytical
formulas are not available.  Even only for the $W^+W^-$ scattering, it
is formidable to convert the numerical results given in
ref.~\cite{Denner:1997kq} to our analysis.\footnote{ By setting
  $m_h=100$~GeV, we reproduced the Born-level $W_LW_L$ scattering
  cross sections shown in ref.~\cite{Denner:1997kq}.  We also
  reproduced by our prescription qualitative behaviors (approximate
  sizes) of the ${\cal O}(\alpha)$ corrections shown there, in the
  region where perturbative convergence holds (entire $\cos\theta$ for
  $\sqrt{s}=200$~GeV, and $\cos\theta\gtrsim 0$ at $\sqrt{s}=1$ and
  5~TeV).  } On the other hand, as we noted in Sec.\,\ref{sec:LO&NLO},
the top-loop contributions are numerically smaller, {\it i.e.}, of the
order of 10\%, as compared to the LO contributions by the singlet
loops.  Hence, the above prescription would be a pragmatic method of
computation for this first study.  We will further discuss this issue
in Sec.\,\ref{sec:Concl+disc}.

The differential cross sections for $W^+_LW^-_L\rightarrow W^+_LW^-_L$
and $W^+_LW^+_L\rightarrow W^+_LW^+_L$ are given by
\begin{align}
\left[\frac{d\sigma}{d\cos\theta}\right]_{W^+_LW^-_L\rightarrow W^+_LW^-_L}
&=
\frac{1}{32\pi s}|{\cal A}_{W^+_LW^-_L\rightarrow W^+_LW^-_L} |^2\,,
\label{eq:W+W-diffxs}
\\
\left[\frac{d\sigma}{d\cos\theta}\right]_{W^+_LW^+_L\rightarrow W^+_LW^+_L}
&=
\frac{1}{64\pi s}|{\cal A}_{W^+_LW^+_L\rightarrow W^+_LW^+_L} |^2\,.
\label{eq:W+W+diffxs}
\end{align}
Here, $\theta$ is the angle between the initial $W^+$ and final $W^+$
momenta in the c.m.\ frame, which satisfies $\cos \theta =
\frac{2t}{s\beta^2}+1$.  We compare the following three cases: (a) SM
tree-level cross section, (b) SM LO cross section, and (c) CSI model
LO cross section, and for the individual cases the amplitudes in
eqs.\,(\ref{eq:W+W-diffxs}) and (\ref{eq:W+W+diffxs}) are given by
\begin{align}
 &{\rm (a):}~ {\cal A}^{\rm SM}_{\rm tree} =
{\cal A}^{\rm quart}+{\cal A}^{\gamma/Z}+
  {\cal A}^{h,{\rm SM}}|_{\rm tree}\, ,
  \label{eq:Amp(a)}
\\
&{\rm (b):}~  {\cal A}^{\rm SM} =
{\cal A}^{\rm quart}+{\cal A}^{\gamma/Z}+
{\cal A}^{h,{\rm SM}}\, ,
\label{eq:Amp(b)}
\\
&{\rm (c):}~ {\cal A}^{\rm CSI} =
{\cal A}^{\rm quart}+{\cal A}^{\gamma/Z}+
{\cal A}^{h} \, .
\label{eq:Amp(c)}
\end{align}
Here the subscript ``$W^+W^-$'' or ``$W^+W^+$'' is suppressed.  
The formulas for the sub-amplitudes ${\cal A}^{\rm
  quart}$, ${\cal A}^{\gamma/Z}$, ${\cal A}^{h,\text{SM}}$ are given
in App.~\ref{app:WWscattering_tree}. In the Higgs-exchange diagrams
only the Higgs propagators are different, {\it i.e.}, the Higgs
propagator is given by
\begin{align}
&{\rm (a):}~\Delta_h^{\text{SM,tree}}=1/(p^2-m_h^2)\,,
\\
&{\rm (b):}~\Delta_h^{\text{SM}}=1/(p^2-m_h^2-\Sigma_h^{t,\text{log}}(p^2))\,,
\\
&{\rm (c):}~\Delta_h^{\text{CSI}}=1/(p^2-m_h^2-\Sigma_h^{t,\text{log}}(p^2)
-\Delta\Sigma_h (p^2))\,,   
\end{align}
where $\Delta\Sigma_h (p^2)$ is defined in eq.\,\eqref{eq:DSigma_h} and
\begin{align}
-\Sigma_h^{t,\text{log}}(p^2)&=-\frac{2N_cy_t^4v^2}{(4\pi)^2}
[B_0(p^2;m_t^2)-B_0(m_h^2;m_t^2)]\,.
\end{align}
It is worth mentioning that close to the pole both $\Delta_h^{\rm
  SM}(p^2)$ and $\Delta_h^{\rm CSI}(p^2)$ behave as
\begin{align}
\Delta_h^{\text{SM,CSI}}(p^2)=\frac{1}{p^2-m_h^2}\times
\Bigl[ 1 + {\cal O}(\xi) \Bigr] +
(\mbox{regular part as $p^2\to m_h^2$}) \,.
\end{align}
Hence, they have the correct pole structure at the LO of $\xi$.

Before showing the numerical results it would be useful to see the
high energy behavior of ${\cal A}^{\rm SM}$ for comparison with the
prediction of the CSI model [{\it c.f.}, eqs.\,\eqref{eq:DampWpWm} and
  \eqref{eq:DampWpWp}]:
\begin{align}
{\cal A}^{\rm SM}_{W^+_LW^-_L\rightarrow W^+_LW^-_L} \approx &
-4\lambda_{\rm H}^{\rm SM}
-\frac{g_Z^2}{2}\left[\frac{s}{t}+\frac{t}{s}+1\right]
\nonumber\\
&
-\frac{2N_cy_t^4}{(4\pi)^2}
\left[ \log \left(\frac{m_t^4}{st}\right) + \text{const.}
\right]
+{\cal O}\left(\frac{m_t^2}{s}\right)\,,
\label{eq:WpWmSMlargeslim}
\\
{\cal A}^{\rm SM}_{W^+_LW^+_L\rightarrow W^+_LW^+_L} \approx &
-4\lambda_{\rm H}^{\rm SM}
-\frac{g_Z^2}{2}\left[\frac{u}{t}+\frac{t}{u}+1\right]
\nonumber\\
&
-
\frac{2N_cy_t^4}{(4\pi)^2}
\left[ \log \left(\frac{m_t^4}{tu}\right) + \text{const.}
\right]
+{\cal O}\left(\frac{m_t^2}{s}\right)\,,
\label{eq:WpWpSMlargeslim}
\end{align}
where $g_Z=\sqrt{g_Y^2+g_2^2}$ [$\,g_Y$ is the gauge coupling of $U(1)_Y$].
The coefficients of the logarithms are consistent
with the $y_t^4$ part of the one-loop beta function of
$\lambda_{\rm H}^{\rm SM}$.\footnote{
The reason why within our prescription
we can ignore the sub-amplitudes ${\cal A}^{t,\text{SM}}$
(defined in the previous sections) is that
there is no diagram with UV divergence proportional to $y_t^4$
therein.
[If we neglect ${\cal A}^{t,\text{SM}}$, the suppressed constants
in eqs.~(\ref{eq:WpWmSMlargeslim}) and (\ref{eq:WpWpSMlargeslim})
are both equal to 4.]
}

\begin{figure}
\begin{center}
  \includegraphics[scale=0.7]{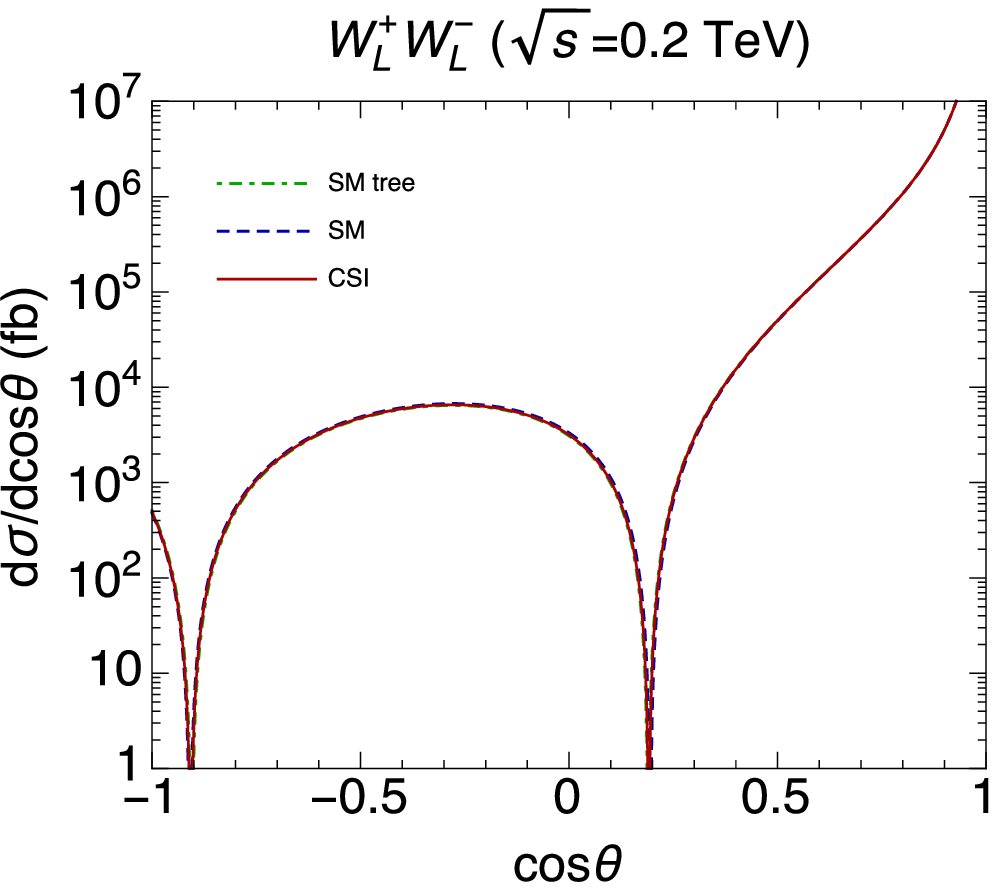}
  \includegraphics[scale=0.7]{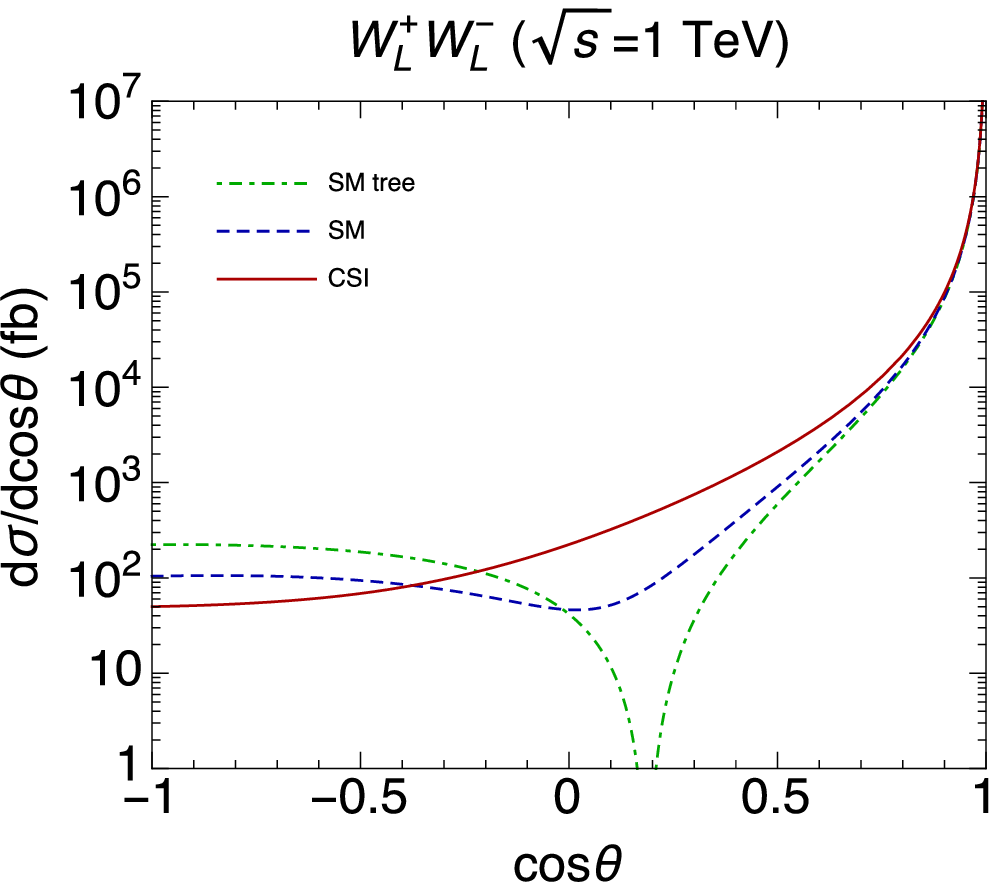}
  \includegraphics[scale=0.7]{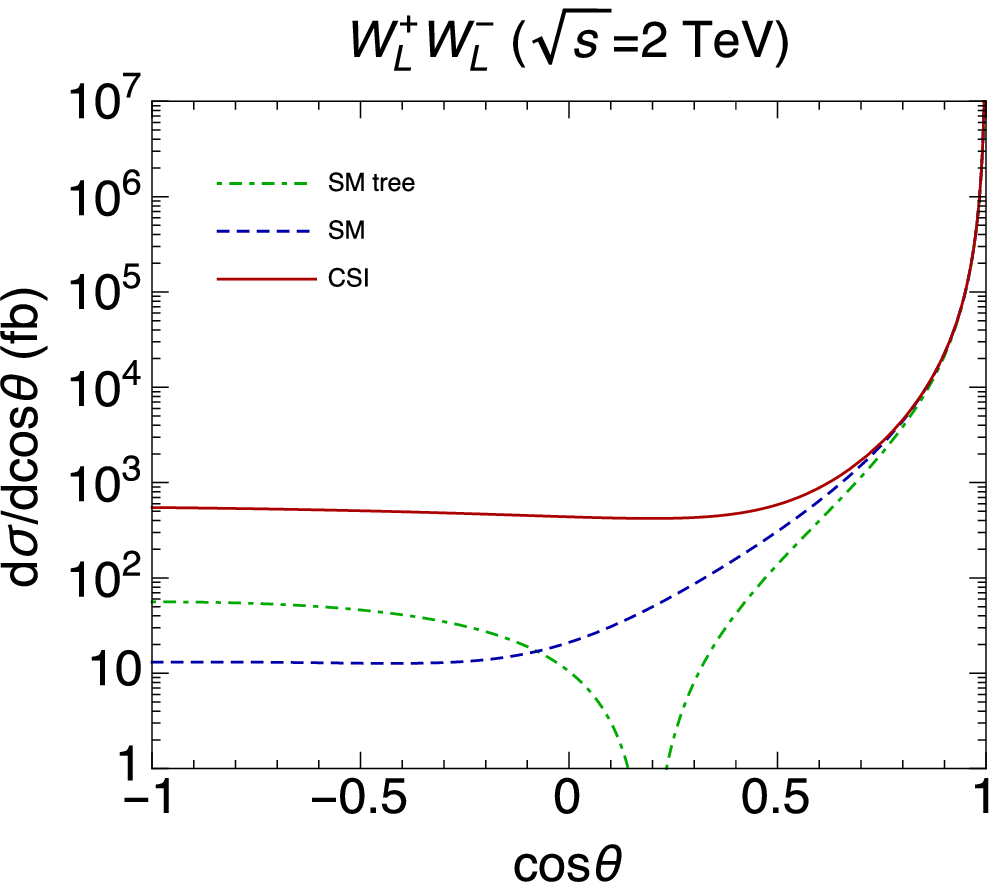}
\end{center}
\caption{\small Differential cross sections for $W^+_LW^-_L\rightarrow
  W^+_LW^-_L$ process for the 
  c.m.\ energies $\sqrt{s}=0.2$~TeV, 1~TeV and 
  2~TeV. We set $N=1$.
  $\theta$
  represents the angle between the initial $W^+$ and final $W^+$
  momenta in the c.m.\ frame. 
  See text for the input parameters.
  Dot-dashed (green) line represents the Born SM cross section,
  dashed (blue) line the SM cross section, and
  solid (red) line the CSI model cross section.
 }
\label{fig:dsigmaWpWm}
\end{figure}

\begin{figure}
\begin{center}
  \includegraphics[scale=0.7]{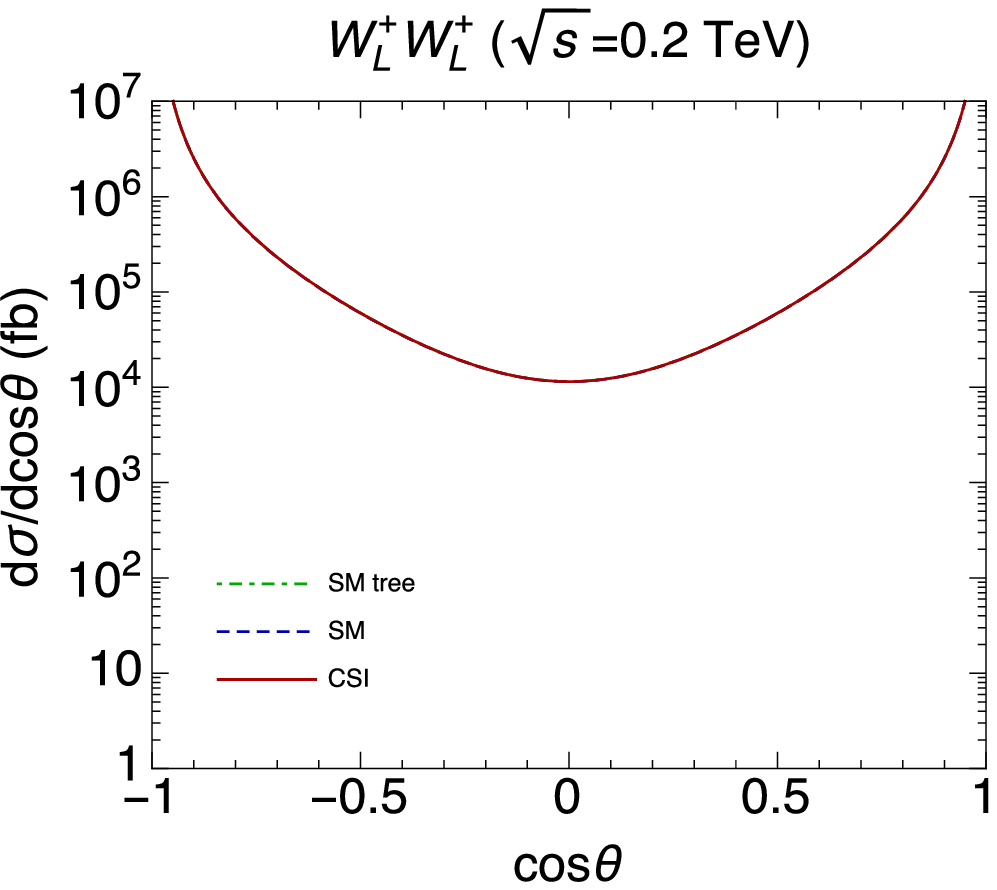}
  \includegraphics[scale=0.7]{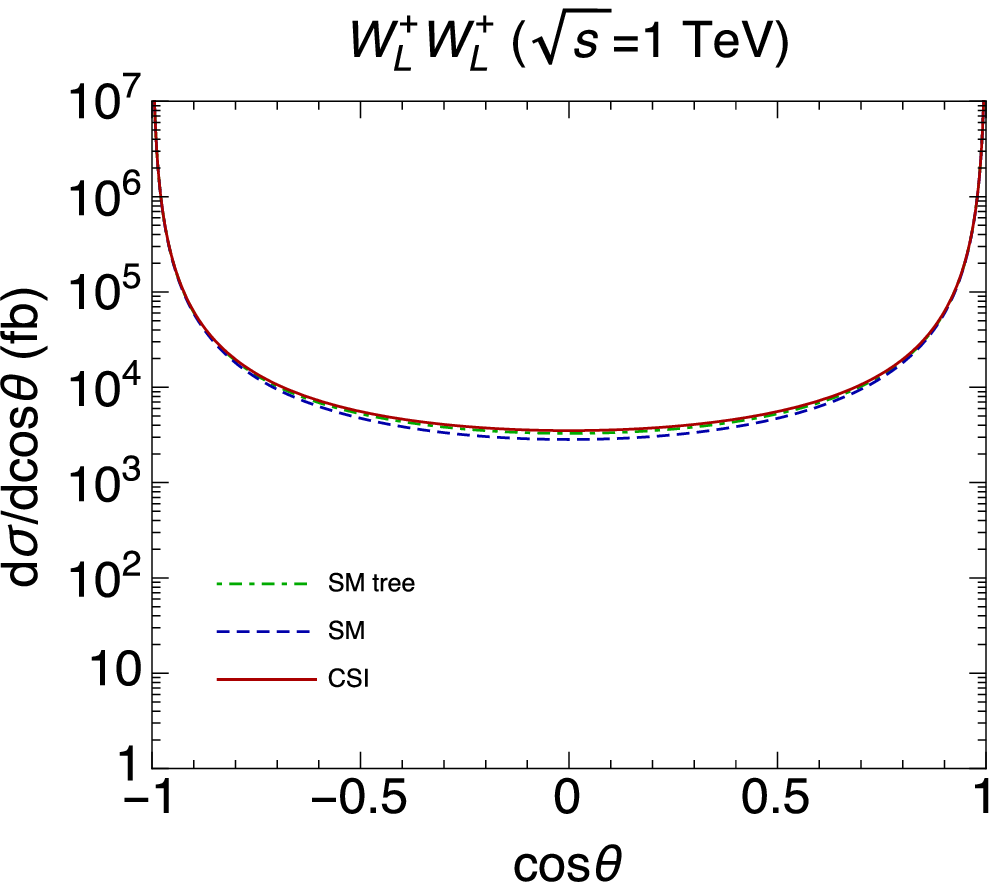}
  \includegraphics[scale=0.7]{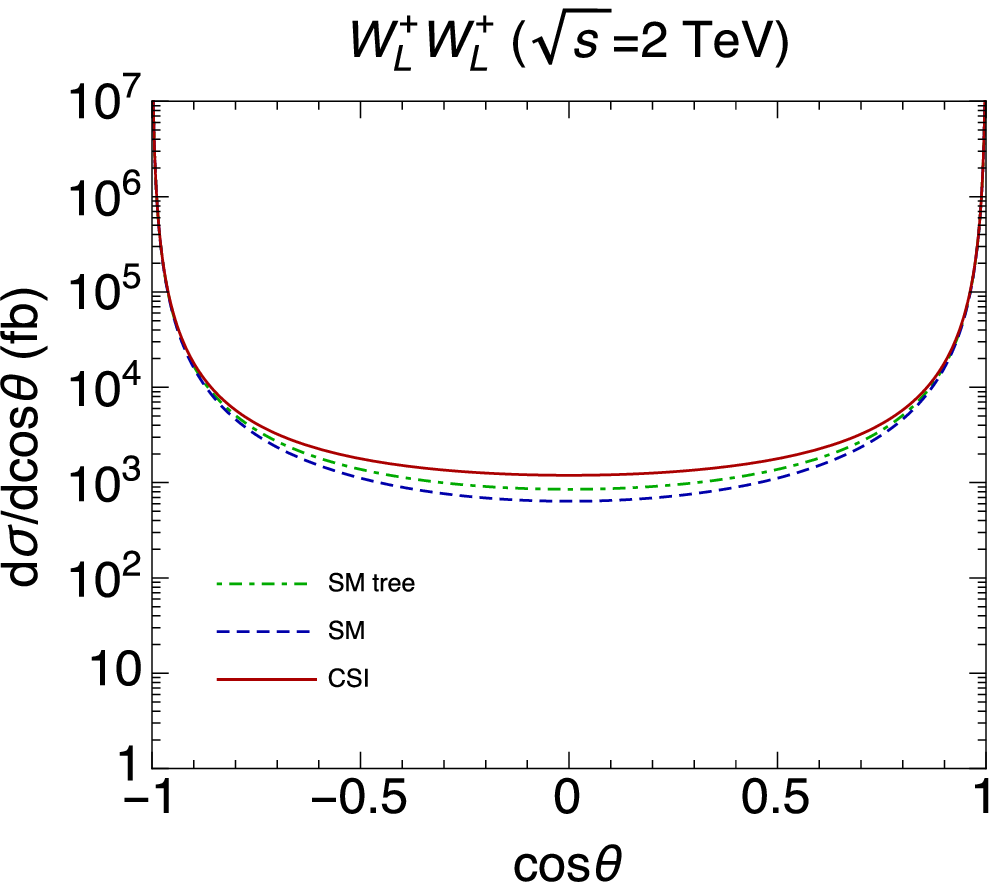}
\end{center}
\caption{\small Same as Fig.\,\ref{fig:dsigmaWpWm} but for
  $W^+_LW^+_L\rightarrow W^+_LW^+_L$ process.}
\label{fig:dsigmaWpWp}
\end{figure}

With the above amplitudes, we compute the cross sections, which are
shown in Fig.\,\ref{fig:dsigmaWpWm} for the $W^+_LW^-_L$ scattering
and in Fig.\,\ref{fig:dsigmaWpWp} for the $W^+_LW^+_L$ scattering.  We
display the case $N=1$ as an example.  The input parameters are taken
as $m_W = 80.385$~GeV, $m_Z = 91.1876$~GeV ($Z$ boson mass), $m_h =
125.03$~GeV, $m_t = 173.34$~GeV, and $g_2 = 0.65178$.  Other
parameters are derived using the tree-level SM relations:\footnote{ It
  is important to maintain these relations, in order to warrant gauge
  cancellation at high energy.  } $\sin^2\theta_W = 1 - m_W^2/m_Z^2$,
$v = 2m_W/g_2$, $y_t = \sqrt{2}m_t/v$.  $\lambda_{\rm HS}$ and $m_s$
are given in Tab.~\ref{table:lambdaHSms}. In $W^+_LW^-_L$ scattering
(Fig.\,\ref{fig:dsigmaWpWm}) we see that the deviation [difference of
  the solid (red) and dashed (blue) lines] is larger at higher energy.
The deviation gets prominent at $\sqrt{s}\gtrsim 1$~TeV.  Note that
the deviation is characteristic to off-shell Higgs bosons as we
discussed below eq.\,(\ref{eq:diff-W+W-}).  For instance, at
$\cos\theta=0.5$, the CSI model cross section is about 2.3\,(1.9)
times larger than the SM cross section at $\sqrt{s}=1\,(2)$~TeV.

\begin{figure}[t]
\begin{center}
  \includegraphics[scale=0.7]{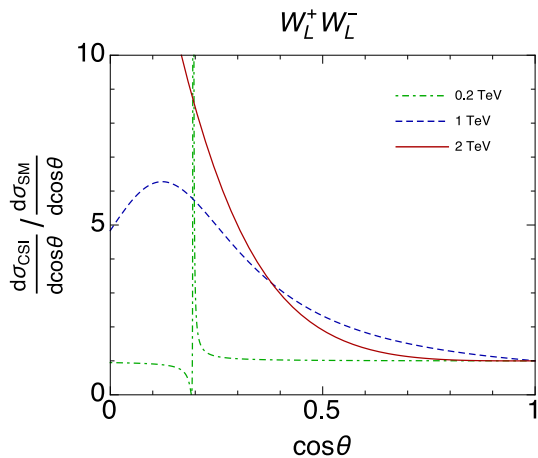}
\end{center}
\caption{\small 
Ratio of the $W^+_LW^-_L$ scattering differential cross
sections for 
the CSI model and SM vs.\ $\cos\theta$,
at $\sqrt{s}=0.2$~TeV (green dot-dashed), 
1~TeV (blue dashed) and 2~TeV (red solid).}
\label{fig:RdsigmaWpWm}
\end{figure}

Nevertheless it might be necessary to observe the deviation at a
smaller angle in order to gain statistics.  Since the deviation
eq.\,(\ref{eq:DampWpWm}) includes an enhancement factor $\sim \log
|t|$ in the forward region, a priori it is not obvious whether the
deviation is highly suppressed in the forward region due to the
enhancement of the SM cross section in that region.\footnote{ The
  cross section exhibits strong enhancement in the forward region
  $\cos\theta \to 1$ due to the $t$-channel gauge and Higgs boson
  exchanges.  At high energy, this can be seen in the term $1/t
  \propto 1/(1-\cos\theta)$ in eq.\,(\ref{eq:WpWmSMlargeslim}).  This
  part is proportional to the gauge coupling $g_Z^2$ and is absent in
  the deviation of the CSI model prediction from the SM prediction,
  eq.\,(\ref{eq:DampWpWm}).  Hence, in general, the deviation can be
  seen more vividly at a larger angle $\theta$, where the cross
  section becomes smaller.  This feature can be seen in the figures.}
In fact the deviation is a complicated function of $s$ and $\theta$,
and can become relevant.  For instance, the CSI model cross section is
larger than the SM cross section by 29\%\,(13\%) at
$\cos\theta=0.8\,(0.9)$ at $\sqrt{s}=1$~TeV.  For comparison, in
Fig.\,\ref{fig:RdsigmaWpWm} we plot the ratio of the differential
cross sections for the CSI model and the SM as a function of
$\cos\theta$ at different c.m.\ energies.

Taken at face value, there is a huge deviation in the backward region
$\cos\theta \lesssim 0$ at high energy as can be seen in
Fig.\,\ref{fig:dsigmaWpWm}.  In this very kinematical region, however,
perturbative convergence of the SM prediction is lost.  This can be
verified by comparing the Born SM cross section and the LO SM cross
section (with only the log-enhanced part of top-loops) in the same
figures.  More accurately, one can confirm this feature in the full
one-loop electroweak corrections computed in
ref.~\cite{Denner:1997kq}.  In this kinematical region we need to
resum certain IR logarithms to stabilize the SM prediction.  Hence,
our predictions for the relative size of the deviation with respect to
the SM cross section are not reliable 
at $\cos\theta\lesssim 0.4$, although the size of the
deviation itself is well under control.

If we increase $N$, the effective coupling $\sqrt{N}\lambda_{\rm HS}$
of the loop correction is unchanged, while the singlet mass $m_s$
becomes smaller.  As a result, the deviation tends to get larger.  On
the other hand, there occurs a cancellation between the singlet
contribution and the SM amplitude in some exceptional kinematical
points, and the deviation becomes small close to such kinematical
points.  For example, in the case $N=4$ and $\cos\theta=0.5$, the CSI
model cross section is about 3.1\,(1.7) times larger than the SM cross
section at $\sqrt{s}=1\,(2)$~TeV.

We can make a similar analysis for the $W^+_LW^+_L$ scattering cross
sections (Fig.\,\ref{fig:dsigmaWpWp}). 
 
In fact, the $W^+_L W^+_L$ scattering channel would be more promising
than the $W^+_L W^-_L$ channel for detecting the deviation from the SM
since the backgrounds, such as $W^+_T W^-_T$ contributions, can be
reduced effectively~\cite{Barger:1990py}.  By definition, the
$W^+_LW^+_L$ differential cross section is symmetric under $\theta \to
\pi-\theta$.  The cross section exhibits strong enhancement in the
forward and backward regions, $\cos\theta \to \pm 1$, due to the $t$-
and $u$-channel gauge and Higgs boson exchanges.  The deviation of the
CSI model prediction from the SM prediction is larger in the central
region $\cos\theta\sim 0$ (where the cross section becomes small) and
at higher energy.  This feature can be seen in the figures.  For
example, at $\cos\theta=0$, the CSI model cross section is
24\%\,(87\%) larger than the SM cross section at
$\sqrt{s}=1\,(2)$~TeV. The deviation may also be important in the
forward or backward region.  The CSI model cross section is larger
than the SM cross section by 9\%\,(25\%) at $\cos\theta=\pm 0.8$ and
by 5\%\,(12\%) at $\cos\theta=\pm 0.9$ when $\sqrt{s}=1\,(2)$~TeV.
Unlike for the $W^+W^-$ scattering, the deviation in the forward
region is larger at $\sqrt{s}=2$~TeV than at $\sqrt{s}=1$~TeV.

\begin{figure}[t]
\begin{center}
  \includegraphics[scale=0.7]{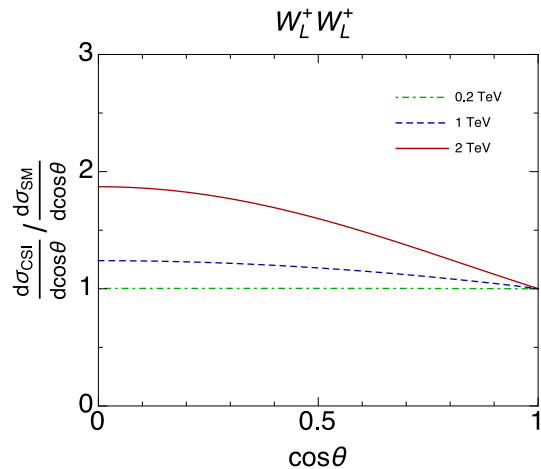}
\end{center}
\caption{\small Same as Fig.\,\ref{fig:RdsigmaWpWm} but for
  $W^+_LW^+_L$ scattering.}
\label{fig:RdsigmaWpWp}
\end{figure}

Convergence of the perturbative prediction of the SM cross section is
good for the $W^+W^+$ scattering.  Thus in the entire range of
$\theta$ and $\sqrt{s}$ analyzed here, we can predict the relative
significance of the deviation reliably.  In
Fig.\,\ref{fig:RdsigmaWpWp} we show the ratio of the differential
cross sections for the CSI model and the SM as a function of $\cos
\theta$, at different c.m.\ energies. If we increase $N$, the
deviation becomes larger.  Differently from the $W^+_LW^-_L$
scattering, there is no cancellation between the singlet contribution
and the SM amplitude.  The deviation for $N=4$ (as an example) becomes
larger than that in $N=1$ case by a factor 1.2--2, depending on
$\cos\theta$ and $\sqrt{s}$ shown in the figures.

For the convenience of the reader, we list values of the scattering
amplitudes at some sample kinematical points in Tab.~\ref{table:amp}.
Note that at $\sqrt{s}>2m_t$, the amplitudes ${\cal A}^\text{SM}$ and
${\cal A}^\text{CSI}$ exhibit imaginary part from the self-energy in
the $s$-channel Higgs propagator for the $W^+W^-$ scattering.

 Before closing this section, we comment on the Landau pole
  and perturbative unitarity.  One may worry about the validity of the
  numerical results in this section due to existence of low-scale Landau
  pole in this model.\footnote{Our definition of the Landau pole
  is the location of the poles of the running coupling constants. }  
  For $N=1$, it
  is located at 3.5--4.7~TeV.  (It becomes higher for larger $N$, {\it
    e.g.}, 16--19~TeV for $N=4$. )
    A more well-defined criterion may be given by the unitarity bounds.\footnote{
    Unitarity should not be violated in all orders of perturbation, as 
long as the Hamiltonian is hermitian, hence perturbative unitarity 
bounds give estimates of the scales where higher-order effects 
become comparable to the lowest-order predictions.
    }
We have
  checked that unitarity of the partial wave amplitudes for the
  $W^+_LW^-_L$ and $W^+_LW^+_L$ scatterings is violated at $\sqrt{s}
  \gtrsim 2.8$~TeV if we substitute the one-loop running coupling
  constants for the renormalized parameters.  Therefore our
  predictions make sense up to slightly below this scale. 
  (In this sense our results at $\sqrt{s}=2$~TeV 
  should be taken with some caution.) On the
  other hand, it should also be stressed that the breakdown of the
  perturbative unitarity originates only from the Landau pole and that
  the theory is well defined perturbatively.  This situation is
  similar to the simple scalar $\phi^4$ theory which has a Landau
  pole.  If we adopt a UV completion of our model, such as in
  ref.~\cite{Dermisek:2013pta}, there is no Landau pole up to Planck
  scale and perturbative unitarity is never violated up to this
  scale.

\begin{table}[t]
\begin{center}
$W^+_LW^-_L$ scattering ($\cos\theta=0.5$)\\
\begin{tabular}{|c|c|c|c|} \hline
$\sqrt{s}$~[TeV]  & $0.2$ & $1$ & 2 \\ \hline
${\cal A}^{\rm CSI}$&
$0.722$ & $0.729-0.105i$ & $0.415 + 0.661i$
\\ \hline    
${\cal A}^{\rm SM}$&
$0.717$ & $0.471- 0.109i$ & $0.553 - 0.114 i$
\\ \hline
${\cal A}^{\rm SM}_{\rm tree}$&
$0.724$ & $0.393$ & $0.379$ \\ \hline
\end{tabular}
\end{center}
\begin{center}
$W^+_LW^+_L$ scattering ($\cos\theta=0$)\\
\begin{tabular}{|c|c|c|c|} \hline
$\sqrt{s}$~[TeV]  & $0.2$ & $1$ & 2 \\ \hline
${\cal A}^{\rm CSI}$&
$-0.487$ & $-1.35$ & $-1.57$
\\ \hline    
${\cal A}^{\rm SM}$&
$-0.486$ & $-1.21$ & $-1.15$
\\ \hline
${\cal A}^{\rm SM}_{\rm tree}$&
$-0.487$ & $-1.30$ & $-1.33$ \\ \hline
\end{tabular}
\caption{\small Values of the amplitudes defined in
  eqs.\,(\ref{eq:Amp(a)}), (\ref{eq:Amp(b)}), (\ref{eq:Amp(c)}) 
for the $N=1$ case. 
$\cos\theta=0.5$ and $\cos\theta=0$ are
chosen for the $W^+_LW^-_L$ and $W^+_LW^+_L$ scattering
processes, respectively.}
\label{table:amp}
\end{center}
\end{table}

\section{Conclusions and discussion}
\label{sec:Concl+disc}
\setcounter{equation}{0}

Although experimental results so far, especially at the LHC, are
almost consistent with predictions of the SM, the mechanism of
electroweak symmetry breaking has not been completely unveiled
yet. Gauge boson scattering is important to understand underlying
physics of electroweak symmetry breaking.

Classical scale invariance with extended Higgs sector is an
alternative scenario for the electroweak symmetry breaking. 
We have computed $WW$ scattering cross sections in a minimal model
with classical scale invariance (CSI model) as a model of new
physics. 
This model
is perturbatively renormalizable and we have developed a theoretical
basis necessary for consistent perturbative computation of Feynman
amplitudes.  This requires a specific assignment of order counting,
which is organized in powers of an auxiliary parameter $\xi$.

The deviation of the CSI model predictions from the SM predictions is
clarified. It arises from the loop correction by singlet field in the
Higgs self-energy and it is incorporated by $\Delta \Sigma_h$, defined
in eq.\,\eqref{eq:DSigma_h}. $\Delta \Sigma_h$ characterizes the
information on symmetry breaking mechanism carried by an off-shell
Higgs boson.\footnote{The definition and the role of $\Delta \Sigma_h$
  are somewhat similar to those of the $S$ parameter of precision
  electroweak corrections, which characterizes information on new
  physics carried by the weak gauge bosons. }  The obtained formulas
can be used for general $W$ polarizations.  We have compared the
scattering amplitudes for the longitudinal $W$ bosons ($W_LW_L\to
W_LW_L$) and NG bosons ($GG\to GG$) and confirmed that they coincide
in the high energy limit, which is consistent with the equivalence
theorem.

The obtained amplitudes for $WW$ scattering enable us to access the
details of the kinematics of the scattering processes, which is
impossible from the effective potential (since it is given by zero
external momentum limit). This point can be seen by looking at the
deviation of the Higgs quartic coupling of the effective
potential:\footnote{ The deviation of the quartic Higgs self-coupling
  for zero external momenta is given by setting $\phi=v$ in
  eq.~(\ref{eq:devquarticcoupl}), which is about three times larger
  than the tree-level SM coupling.  This is consistent with the
  estimates in \cite{Dermisek:2013pta,Endo:2015ifa}.  }
\begin{align}
\frac{1}{3!}\frac{\partial^4}{\partial \phi^4}&
[V_\text{eff}^\text{CSI}(\phi)-V_\text{eff}^\text{SM}(\phi)] 
\nonumber\\
&=\frac{1}{3!}
\frac{\partial^4}{\partial \phi^4}	
\left[
\frac{1}{4}\{(\lambda_\text{H}+\delta\lambda_\text{H})
  -(\lambda_\text{H}^\text{SM}+\delta\lambda_\text{H}^\text{SM})\}\phi^4
+\frac{N\lambda_\text{HS}^2}{4(4\pi)^2}\phi^4
\left(
\log\frac{\lambda_\text{HS}\phi^2}{\mu^2}
-\frac{3}{2}\right)
\right]
\nonumber\\
&=
\frac{N\lambda_\text{HS}^2}{(4\pi)^2}
\left[\log\frac{\lambda_{\rm HS}\phi^2}{m_s^2}
+\frac{8}{3}\right]
\label{eq:devquarticcoupl}
\,  .
\end{align}
This should be compared with 
\begin{align}
-\frac{1}{4}\Delta {\cal A}_{W^+_LW^-_L\to W^+_LW^-_L}~\to~ &
\frac{N\lambda_{\rm HS}^2}{(4\pi)^2}\,
\left[
\log \left(\frac{\sqrt{st}}{m_s^2} \right)
-2 \right]\, ,
\\
-\frac{1}{4}\Delta {\cal A}_{W^+_LW^+_L\to W^+_LW^+_L}~\to~ &
\frac{N\lambda_{\rm HS}^2}{(4\pi)^2}\,
\left[
\log \left(\frac{\sqrt{tu}}{m_s^2} \right)
-2 \right]\, ,
\end{align}
obtained from eqs.\,(\ref{eq:DampWpWm}) and (\ref{eq:DampWpWp}).
Comparing them, one could expect the anomalous $\frac{N\lambda_{\rm
    HS}^2}{(4\pi)^2}\log s$ behavior of ${\cal A}_{W_LW_L\to W_LW_L}$
in the high energy limit from the effective potential if $\phi$ is
interpreted as $\sqrt{s}$. However, it is impossible to give the other
terms correctly or make predictions for lower energy scattering from
the effective potential. From this viewpoint the computation based on
the proper order counting using the auxiliary parameter $\xi$ is
crucial for the accurate predictions for $WW$ scattering processes.

For $W^+_LW^-_L$ scattering we predict +29\%\,(+13\%) deviation at,
{\it e.g.}, $\cos\theta=0.8\,(0.9)$ at $\sqrt{s}=1$~TeV with $N=1$.
($\theta$ is the angle between the incident and scattered $W^+$
bosons, and the cross section is expected to increase as
$\cos\theta\to 1$.)  For $W^+_LW^+_L$ scattering we may profit from a
larger cross section around $\cos\theta \simeq \pm 1$, and a deviation of
+25\%\,(+12\%) at $\cos\theta=\pm 0.8\,(\pm 0.9)$ and $\sqrt{s}=2$~TeV
is predicted.  If we increase $N$, the deviation tends to become
larger for both cases.

In summary we can describe the characteristic aspects of the CSI model
as follows.  (1) The deviations in $W_LW_L$ cross sections are large,
and (2) they can be quantified by well-known loop functions.

Finally some remarks for future studies are in order.  Our main
purpose is to set up a theoretical basis for implementing the
predictions of the CSI model to Monte Carlo event generators.  We have
found a simple prescription to modify the SM predictions, as stated
above.  This prescription is valid also for off-shell $W$ processes as
it is clear from the derivation.  In addition, it is independent of
gauge choice for the electroweak gauge symmetry since the portal
interaction is not affected by the gauge fixing condition.  As we
checked in Secs.~\ref{sec:W+W-amp} and \ref{sec:W+W+amp}, it preserves
gauge cancellation and satisfies the equivalence theorem
\cite{Cornwall:1974km,Vayonakis:1976vz,Chanowitz:1985hj} for the CSI
model.  Therefore, the prescription is suited for implementation to
Monte Carlo simulation for collider experiments.

It is not trivial whether the model can be tested using $W_LW_L$
scattering processes at the LHC experiments.  According to
\cite{Barger:1990py}, luminosity of initial $W_L$s would not be too
suppressed compared to that of $W_T$s.  Past researches, such as
refs.~\cite{Barger:1990py,Dicus:1990fz}, or recent
works~\cite{Han:2009em,Doroba:2012pd,Campbell:2015vwa}, would be useful
for devising kinematical cuts to enhance signal to background ratio in
collider searches. Use of $\tau$ final states may help to
enhance $W_L$ signals. Detailed study will be given
elsewhere~\cite{InProgress}.

Clearly it is important to have accurate predictions of the SM
predictions for $WW$ scattering processes at the LHC experiments.  Up
to now, perturbative QCD corrections are available at the NLO and 
next-to-next-to-leading order 
for various kinematical distributions.  Full NLO QCD corrections are
implemented in Monte Carlo event generators.  On the other hand, full
NLO electroweak corrections have not been implemented in event
generators so far, despite extensive efforts in this direction.
(See, {\it e.g.}, ref.~\cite{Gieseke:2014gka} and references therein.) 

Among the SM electroweak corrections, phenomenologically electroweak
Sudakov logarithms \cite{Fadin:1999bq} are known to be important in
the processes involving high energy $W$ bosons.  (See, {\it e.g.},
ref.~\cite{Kuhn:2011mh} and references therein.)  We have not
incorporated these effects accurately in this study. 
They will be taken into account carefully when we make a more realistic 
testability study.
We note that, as far as the deviations of the CSI model predictions from
the SM predictions are concerned, Sudakov logarithms are irrelevant,
so that it does not affect the prescription which we propose.

\section*{Acknowledgement}
The work of Y.S. was supported in part by Grant-in-Aid for scientific
research No. 23540281 from MEXT, Japan.

\appendix
\section*{Appendices}

We collect details of the argument and formulas.  In
App.\,\ref{app:expparam}, we show the effective expansion parameter of
the CSI model and the SM with our specific order counting.  In
App.\,\ref{app:loopfunc}, loop functions are defined.  In
App.\,\ref{app:WWscattering_tree}, sub-amplitudes for $W_LW_L\to
W_LW_L$ processes are given analytically.  In
App.\,\ref{ap:GGscattering}, NG boson scattering amplitudes are
computed.

\section{Effective expansion parameter}
\label{app:expparam}

In this appendix we explain the details of the expansion in terms of
the parameter $\xi$.  We show that the effective expansion parameter
is given by eq.\,(\ref{eq:ord-xi}) if we rescale the couplings by
eq.\,(\ref{eq:coupling-orders}) and expand in $\xi$.

Before the discussion, we note that we are particularly interested in
the high energy limit of Feynman amplitudes. For instance, in the $W$
boson scattering cases which we have mainly discussed, we organize a
Feynman amplitude in series expansion in the inverse of the
c.m.\ energy, $A(s) = a_0 s^0 + a_1 s^{-1} + a_2 s^{-2} + \cdots$, due
to existence of gauge cancellation.  (External lines are taken to be
on-shell, and other kinematical parameters such as $s/t$ are taken to
be order one.) Then we expand each $a_n$ in terms of $\xi$.  In the
following argument we suppress $s$ for simplicity.  We note that the
following argument is valid also in the case that external momentum
invariants $p_i\!\cdot\! p_j$ are set equal to either $m_h^2$, $m_s^2$
or zero, which correspond to the computations in Sec.~\ref{sec:setup}.

Let us begin by making a consistency check using
eqs.\,(\ref{eq:tadpole^CSI}), (\ref{eq:Sigma_h^CSI}),
(\ref{eq:mhapp}).  In these equations, $\lambda_{\rm H}$,
$\lambda_{\rm HS}^2/(4\pi)^2$ and $y_t^4/(4\pi)^2[\propto y_t^2
  m_t^2/(4\pi)^2]$ are treated as the same order quantities, if we
take into account the loop factors as well.  It is equivalent to
treating $\lambda_{\rm H}/(4\pi)^2$, $\lambda_{\rm HS}^2/(4\pi)^4$ and
$y_t^4/(4\pi)^4$ as the same order quantities, which is consistent
with eq. (2.9).  Thus, in the LO analysis in Sec. 2.3, we correctly
compare the quantities formally counted as the same order.  It is true
for the SM as well.

Let us consider the effective action of the CSI model, which is the
generating functional of the amputated one-particle irreducible (1PI)
Green functions:
\begin{align}
\Gamma[\varphi(x)]=\sum_n
\int d^Dx_1\cdots d^Dx_n\,
\Gamma^{(n)}_{i_1,\cdots,i_n}(x_1,\cdots,x_n)
\,\varphi_{i_1}(x_1)\cdots \varphi_{i_n}(x_n) .
\end{align}
Here, we consider the effective action in the symmetric phase, and
$\varphi$ denotes the collection of all the fields in the model, {\it
  e.g.}, $\varphi_1=H$, $\varphi_2=S_i$, $\varphi_3=W_\mu^a$,
$\varphi_4=\psi_L$, etc., but the details are irrelevant in the
following argument.  Hence, $\Gamma[\varphi(x)]$ is invariant under
the SM gauge transformation and the global $O(N)$ transformation
$\varphi\to G_{\text{SM}\times O(N)}\,\varphi$.

For simplicity, we concentrate on $\lambda_{\rm H}$
and $\lambda_{\rm HS}$ and neglect all the other couplings.
Before we make the rescaling eq.\,(\ref{eq:coupling-orders}),
the perturbative expansion of each 
1PI Green function takes a form
\begin{align}
\Gamma^{(n)}_{i_1,\cdots,i_n}(x_1,\cdots,x_n)
= & \,
\lambda_{\rm H}^{a(n;i_1,\cdots,i_n)}
\lambda_{\rm HS}^{b(n;i_1,\cdots,i_n)}
\nonumber\\
&
\times \sum_{k,m=0}^\infty 
\left[ \frac{\lambda_{\rm H}}{(4\pi)^2}\right]^k
\left[ \frac{\lambda_{\rm HS}}{(4\pi)^2}\right]^m
\Gamma^{(n,k,m)}_{i_1,\cdots,i_n}(x_1,\cdots,x_n) ,
\end{align}
where the powers of $\lambda_{\rm H}$ and $\lambda_{\rm HS}$
corresponding to the tree-level 1PI Green function are factored
out.\footnote{ If there are more than one combination of $(a,b)$,
  which contribute to the tree-level 1PI Green function, we take the
  sum over all the combinations.  } Namely,
$\Gamma^{(n,0,0)}_{i_1,\cdots,i_n}$ denotes the tree-level 1PI Green
function for the particles $\varphi_{i_1},\cdots,\varphi_{i_n}$, while
for $k+m>0$, $k+m$ is equal to the number of loops.\footnote{ It is
  understood that $\Gamma_{i_1,\cdots,i_n}^{(n,k,m)}$ is expanded 
  in $1/s$ after Fourier transformation.}

After the rescaling eq.\,(\ref{eq:coupling-orders}),
we have
\begin{align}
\Gamma^{(n)}_{i_1,\cdots,i_n}(x_1,\cdots,x_n)
& \, \to \xi^{2a+b}
\lambda_{\rm H}^{a}
\lambda_{\rm HS}^{b}
\nonumber\\
&
\times \sum_{k,m=0}^\infty \xi^{2k+m}
\left[ \frac{\lambda_{\rm H}^{1/2}}{(4\pi)}\right]^{2k}
\left[ \frac{\lambda_{\rm HS}}{(4\pi)^2}\right]^m
\Gamma^{(n,k,m)}_{i_1,\cdots,i_n}(x_1,\cdots,x_n) .
\end{align}
Thus, for each power of $\lambda_{\rm H}^{1/2}/(4\pi)$
or $\lambda_{\rm HS}/(4\pi)^2$, the power of $\xi$ is raised by one.
This is consistent with eq.\,(\ref{eq:ord-xi}).

In the symmetry broken phase, we replace the fields as $\varphi_i \to
v_i + \delta \varphi_i$, where $v_i$ is the VEV of $\varphi_i$.  Then
we re-expand the effective action in $\delta \varphi_i$, and the
expansion coefficients represent the 1PI Green functions of the broken
phase.  Since we do not assign powers of $\xi$ to $v_i$, the relation
between the order counting in $\xi$ and in $\lambda_{\rm
  H}^{1/2}/(4\pi)$ and $\lambda_{\rm HS}/(4\pi)^2$ is unchanged from
the symmetric phase.

\begin{figure}[t]
\begin{center}
  \includegraphics[scale=0.5]{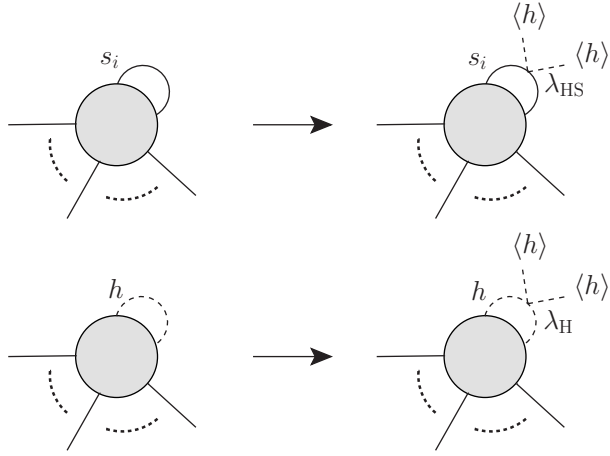}
\end{center}
\caption{\small Insertion of $\lambda_{\rm HS} v^2 s_is_i$ vertex to a
  1PI diagram, and insertion of $\lambda_{\rm H} v^2 hh$ vertex to a 1PI
  diagram, respectively.  These manipulations do not increase the
  number of loops.  }
\label{fig:ordcounting}
\end{figure}

The different feature in the broken phase is that one can increase the
number of vertices in 1PI diagrams without increasing the number of
external legs.  For instance, as shown in Fig.\,\ref{fig:ordcounting},
we can insert one $\xi\lambda_{\rm HS}$ vertex along the internal
singlet line or insert one $\xi^2\lambda_{\rm H}$ vertex along the
internal Higgs line, together with $v^2$.  This manipulation does not
increase the number of loops, hence the power of $1/(4\pi)^2$ does not
change.  However, the additional dimensionful parameter $v^2$ should
be compensated by some dimensionful parameter in the denominator,
which appears as a result of loop integrals.  According to our
assumption for the external momentum invariants, this compensation
factor, combined with the $v^2$, should have a form\footnote{
Since kinematical parameters are not
accompanied by powers of $\xi$, powers of $\xi$ appear only from the
couplings and particles' masses, as listed in Tab.~\ref{table:OC}.}
\begin{align}
C=\frac{v^2}{\xi m_s^2} \sum_{n} c_n \left(
\frac{\xi m_h^2}{m_s^2}\right)^n \,.
\end{align}
Note that
\begin{align}
&
\frac{\xi m_h^2}{m_s^2}=\xi \frac{\lambda_{\rm HS}}{(4\pi)^2}
\times \text{const.}
+{\cal O}(\xi^2) \,,
\label{eq:mh^2/ms^2}
\\ &
\frac{v^2}{\xi m_s^2}=\frac{1}{\xi \lambda_{\rm HS}}
\times \text{const.}
+{\cal O}(\xi^0) \,,
\label{eq:v^2/ms^2}
\end{align}
by eqs.\,(\ref{eq:mhapp}) and (\ref{eq:ms}).  If we multiply $C$ with
the inserted vertex $\xi \lambda_{\rm HS}$ or $\xi^2 \lambda_{\rm H} =
\xi \lambda_{\rm HS} \times [\xi \lambda_{\rm H}^{1/2}/(4\pi)]^2\times
    [\xi \lambda_{\rm HS}/(4\pi)^2]^{-1}$, we obtain a series
    expansion, where $\xi$ is accompanied by each power of
    $\lambda_{\rm H}^{1/2}/(4\pi)$ or $\lambda_{\rm HS}/(4\pi)^2$.  In
    a similar manner, one can verify that inclusion of the VEV of the
    Higgs field does not change the order counting.  Higher powers of
    $\xi$ on the right-hand side of eqs.\,(\ref{eq:mh^2/ms^2}) and
    (\ref{eq:v^2/ms^2}) can be determined iteratively by applying the
    above method using the terms already determined at lower orders.

By way of example, the one-loop diagram and 
counterterm in
Fig.\,\ref{fig:SingletSelfenergy} contribute to the NLO correction
to the $\xi$ expansion of the
singlet mass as [see eq.\,(\ref{eq:ms})]
\begin{align}
\xi m_s^2&=\xi \lambda_{\rm HS} v^2 + \xi^2 
\frac{\lambda_{\rm HS}^2 v^2}{(4\pi)^2}\times \text{const.} 
\nonumber \\
&=\xi \lambda_{\rm HS} v^2 \left[ 1+ \xi 
\frac{\lambda_{\rm HS}}{(4\pi)^2}\times \text{const.} \right] .
\end{align}
In the language of the above effective action, the LO term corresponds
to $(a,b,k,m) = (0,1,0,0)$, hence $\xi^{2a+b+2k+m} = \xi^1$; the NLO
term corresponds to $(a,b,k,m) = (0,1,0,1)$, hence $\xi^{2a+b+2k+m} =
\xi^2$. We can apply this argument including $y_t$ or to the case of
the SM.

\section{Loop functions}
\label{app:loopfunc}
\setcounter{equation}{0}

We give loop functions in $D=4-2\epsilon$ dimension introducing the
renormalization scale $\mu$:
\begin{align}
&\mu^{2\epsilon} \int \frac{d^Dq}{(2\pi)^D}\frac{1}{q^2-m^2} =
\frac{i}{(4\pi)^2}m^2\tilde{A}_0(m^2)\,,
\\
&\mu^{2\epsilon} \int \frac{d^Dq}{(2\pi)^D}\frac{1}{(q^2-m^2_1)((q+p)^2-m^2_2)}
=\frac{i}{(4\pi)^2}B_0(p^2;m^2_1,m^2_2)\,,
\\
&\mu^{2\epsilon} \int \frac{d^Dq}{(2\pi)^D}\frac{q_\mu}{(q^2-m^2_1)((q+p)^2-m^2_2)}
=\frac{i}{(4\pi)^2}p_\mu B_1(p^2;m^2_1,m^2_2)\, .
\end{align}
The following expressions are sufficient for our discussion:
\begin{align}
 &\tilde{A}_0(m^2) =
  \frac{1}{\bar{\epsilon}}+\log \left(\frac{\mu^2}{m^2}\right)+1 \,,
  \\
 & B_0(p^2;m_1^2,m_2^2)
  =\frac{1}{\bar{\epsilon}}
  +\log\mu^2
  -\int_0^1 dx \, \log[\, (1-x)m_1^2 + xm_2^2 - x(1-x)p^2 -i0\,]\,,
  \\
  &B_1(p^2;m^2,m^2)=-\frac{1}{2}B_0(p^2;m^2,m^2)\,,
\end{align}
with $1/\bar{\epsilon}=1/\epsilon -\gamma+\log 4\pi$ ($\gamma$ is
Euler's number) and
\begin{align}
B_0(p^2;m^2)-B_0(q^2;m^2)
&=
f(q^2/m^2)-f(p^2/m^2) \, ,
\end{align}
where
\begin{align}
f(z)&=
2+\int_0^1 dx \, \log[\, 1 - x(1-x)z - i0 \,]
\rule[-7mm]{0mm}{15mm}
\nonumber\\
&=
\left\{
\begin{array}{llc}
\displaystyle
\displaystyle
 2 \sqrt{-d}\,\arctan\biggl(\frac{1}{\sqrt{-d}}\biggr) \, ,
&~~~& 0<z<4
\\
~~~~~~
2 \,,&& z=0
\\
\displaystyle
 \sqrt{d}\,\biggl[
\log\,\biggl| \frac{1+\sqrt{d}}{1-\sqrt{d}}\biggr| -\pi i \,
\theta(z-4) \biggr] \, ,
&~~~& z<0~~\mbox{or}~~ z\geq 4
\end{array}
\right. \, ,
\label{eq:def_f(z)}
\end{align}
with $d=1-4/z$ and $\theta(x)$ is the step function which is equal to
one if $x\ge 0$ and zero if $x<0$.    In the text we
express $B_{i}(p^2;m^2,m^2)$ as $B_{i}(p^2;m^2)$ ($i=0,\,1$) for
simplicity.

Asymptotic expressions of $B_0(p^2;m^2)$ in the limit  $|p^2|\gg
m^2$ or $|p^2| \ll m^2$ are useful:
\begin{align}
  &B_0(p^2;m^2) \xrightarrow{|p^2|\gg m^2}
  \frac{1}{\bar{\epsilon}} + \log \left(\frac{\mu^2}{-p^2}\right)+2
  +{\cal O}\left(\frac{m^2}{|p^2|}\right)\,,
  \\
  &B_0(p^2;m^2) \xrightarrow{|p^2|\ll m^2}
  \frac{1}{\bar{\epsilon}} + \log \left(\frac{\mu^2}{m^2}\right)
  +{\cal O}\left(\frac{|p^2|}{m^2}\right)\,,
\end{align}
and for $B_0(m^2;m^2,m'^2)$,
\begin{align}
B_0(m^2;m^2,m'^2) \xrightarrow{m^2\gg m'^2}
\frac{1}{\bar{\epsilon}} + \log \left(\frac{\mu^2}{m^2}\right)+2
  +{\cal O}\left(\frac{m'^2}{m^2}\right)\,.
\end{align}

\section{$W$ boson scattering (in the SM)}
\label{app:WWscattering_tree}
\setcounter{equation}{0}

Tree-level amplitudes for longitudinally-polarized $W$ boson
scattering processes in the standard model are given. The amplitudes
for the diagrams (i), (ii) and (iii) in Figs.\,\ref{fig:AmpWpWm} and
\ref{fig:AmpWpWp} are given by
\begin{align}
{\cal A}_{W^+_LW^-_L}^{\rm quart} &=
\frac{g_2^2}{4m_W^4\beta^4}(t^2+st\beta^2+3st\beta^4+s^2\beta^6)\,,
\\
{\cal A}_{W^+_LW^-_L}^{\gamma/Z} &=
\frac{g_2^2s^2}{16m_W^4}
\left[\frac{\sin^2\theta_W}{s}+\frac{\cos^2 \theta_W}{s-m_Z^2}\right]
\nonumber \\ &~~~
\times
\left\{-17t+u+2(-4s+7t+u)\beta^2+(8s-t+u)\beta^4\right\}
\nonumber \\ &
+
\frac{g_2^2}{16m_W^4\beta^4}
\left[\frac{\sin^2\theta_W}{t}+\frac{\cos^2 \theta_W}{t-m_Z^2}\right]
\nonumber \\ &~~~
\times
\bigl\{
4t^2u-s^3\beta^4(1-\beta^2)^2
+s^2\beta^2(\beta^2-1)(4t-(16t+u)\beta^2+u\beta^4)
\nonumber \\ &~~~~~
-4st(t-(6t+u)\beta^2+(6t+u)\beta^4)
\bigr\}\,,
\\
{\cal A}_{W^+_LW^-_L}^{h,{\rm SM}} &=
-\frac{g_2^2}{16m_W^2\beta^4}
\left[\frac{s^2(1+\beta^2)^2\beta^4}{s-
  {({m^{\rm SM}_{h,\text{tree}}})}^2
  - \Sigma_h^{\rm SM}(s) }
  +\frac{\left\{2t+s\beta^2(1-\beta^2)\right\}^2}{t- 
{({m^{\rm SM}_{h,\text{tree}}})}^2
  - \Sigma_h^{\rm SM}(t) }
  \right]\,,
  \nonumber \\
\end{align}
 for $W^+_LW^-_L \rightarrow W^+_L W^-_L$ and
\begin{align}
{\cal A}_{W^+_LW^+_L}^{\rm quart} &=
\frac{g_2^2}{4m_W^4\beta^4}(-2t^2-2st\beta^2+s^2\beta^6)\,,
\\
{\cal A}_{W^+_LW^+_L}^{\gamma/Z} &=
\frac{g_2^2}{16m_W^4\beta^4}
\left[\frac{\sin^2\theta_W}{t}+\frac{\cos^2 \theta_W}{t-m_Z^2}\right]
\nonumber \\ &~~~
\times
\bigl\{
-4t^2u+s^3(\beta^2-1)^2 \beta^4
-s^2\beta^2(\beta^2-1)(4t-(16t+u)\beta^2+u\beta^4)
\nonumber \\ &~~~~
+4st(t-(6t+u)\beta^2+ (6t+u)\beta^4)
\bigr\}
\nonumber \\ &
+
\frac{g_2^2}{16m_W^4\beta^4}
\left[\frac{\sin^2\theta_W}{u}+\frac{\cos^2 \theta_W}{u-m_Z^2}\right]
\nonumber \\ &~~~
\times
\bigl\{
-4t^3+s^3\beta^4(1-3\beta^2)^2+4st^2(1-7\beta^2+5\beta^4)
\nonumber \\ &~~~~
-s^2t\beta^2(-4+29\beta^2-30\beta^4+\beta^6)
\bigr\}\,,
\\
{\cal A}_{W^+_LW^+_L}^{h,{\rm SM}} &=
-\frac{g_2^2}{16m_W^2\beta^4}
\left[\frac{\left\{2t+s\beta^2(1-\beta^2)\right\}^2}{t- 
{({m^{\rm SM}_{h,\text{tree}}})}^2
  - \Sigma_h^{\rm SM}(t) }
  +\frac{\left\{2t+s\beta^2(1+\beta^2)\right\}^2}{u- 
{({m^{\rm SM}_{h,\text{tree}}})}^2
  - \Sigma_h^{\rm SM}(u) }
  \right]\,,
\end{align}  
for $W^+_LW^+_L \rightarrow W^+_L W^+_L$.  These are consistent with
the results in refs.~\cite{Duncan:1985vj,Barger:1990py,Denner:1997kq} 

\section{Nambu-Goldstone boson scattering}
\label{ap:GGscattering}
\setcounter{equation}{0}

In this section we give the amplitudes for charged NG boson
scatterings.\footnote{ We neglect the contributions of top-quark loops
  for simplicity.  In particular they cancel in the differences of the
  CSI model and the SM predictions, eqs.~(\ref{eq:DAmpGpGm}) and
  (\ref{eq:DAmpGpGp}) } According to the equivalence theorem
\cite{Cornwall:1974km,Vayonakis:1976vz,Chanowitz:1985hj}, the
amplitude for $W^+W^-\rightarrow W^+W^-$ ($W^+W^+ \rightarrow W^+W^+$)
should agree with that for $G^+G^-\rightarrow G^+G^-$
($G^+G^+\rightarrow G^+G^+$) in the high energy limit. Thus the
amplitudes for NG boson scattering can be used for checking the high
energy behaviors of the $W$ boson scattering amplitudes, which is
especially important to examine the deviations from the SM predictions.

For computation it is useful to rewrite the Higgs quartic coupling and
its counterterm in the SM using eqs.\,\eqref{eq:tadpole^CSI},
\eqref{eq:tadpole^SM}, \eqref{eq:mh^CSI}, and \eqref{eq:mh^SM}:
\begin{align}  
\lambda_{\rm H}^{\rm SM}+\delta\lambda_{\rm H}^{\rm SM} &=
-\frac{N_cy_t^4}{(4\pi)^2}\tilde{A}_0(m_t^2)
-\frac{N\lambda_{\rm HS}^2}{(4\pi)^2}
\left[B_0(m_h^2;m_s^2)-\tilde{A}_0(m_s^2)\right]
+{\cal}O(\xi^3)\,.
\label{eq:lambdaH^SM}
\end{align}
Combining with eq.\,\eqref{eq:tadpole^CSI}, we obtain
\begin{eqnarray}
  (\lambda_{\rm H}+\delta\lambda_{\rm H})
  -(\lambda_{\rm H}^{\rm SM}+\delta\lambda_{\rm H}^{\rm SM} ) =
  \frac{N\lambda_{\rm HS}^2}{(4\pi)^2}B_0(m_h^2;m_s^2)
  +{\cal O}(\xi^3)\,.
  \label{eq:Dlambda_H}
\end{eqnarray}

\subsection{$G^+G^-\rightarrow G^+G^-$ scattering}

\begin{figure}[t]
\begin{center}
 \includegraphics[scale=0.8]{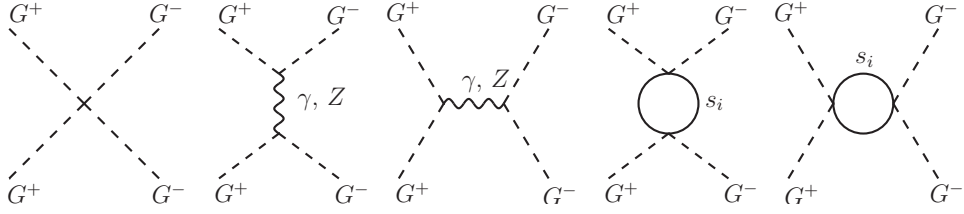}
\end{center}
\caption{\small Diagrams for $G^+G^-\rightarrow G^+G^-$
  scattering. (Time flows upwards.) The last two diagrams are the new
  contributions in the CSI model.}
\label{fig:AmpGpGm}
\end{figure}

We derive the amplitude for $G^+G^-\rightarrow G^+G^-$ scattering in
Landau gauge, which is drawn in Fig.~\ref{fig:AmpGpGm}. 
Crucial differences from the SM amplitude reside in the following two
points. {\it (a)} The Higgs quartic coupling is different (at tree
level). {\it (b)} The singlet-loop diagram gives non-negligible
corrections.  We assign momentum of each particle as
$G^+(p_1)G^-(p_2)\rightarrow G^+(k_1)G^-(k_2)$, and the results are
given by
\begin{align}
  &{\rm CSI}:~~{\cal A}^{\rm CSI}_{G^+G^- \rightarrow G^+G^-} =
  {\cal A}^{\rm quart}_{G^+G^-}+{\cal A}^{\gamma/Z}_{G^+G^-}+{\cal A}^{s}_{G^+G^-}\,,
   \nonumber \\
   &{\rm SM}:~~{\cal A}_{G^+G^- \rightarrow G^+G^-}^{\rm SM} =
   {\cal A}^{\rm quart,SM}_{G^+G^-}+{\cal A}^{\gamma/Z}_{G^+G^-} \,,
\end{align}
where ${\cal A}^{\rm quart}_{G^+G^-}$, ${\cal A}^{\rm quart,SM}_{G^+G^-}$ are
the quartic vertex, given as
\begin{align}
  {\cal A}^{\rm quart}_{G^+G^-} &=
  -4(\lambda_{\rm H}+\delta \lambda_{\rm H})\,,
  \nonumber \\
  {\cal A}^{\rm quart,SM}_{G^+G^-} &=
  -4(\lambda_{\rm H}^{\rm SM}+\delta \lambda_{\rm H}^{\rm SM})\,,
\end{align}
and ${\cal A}^{\gamma/Z}_{G^+G^-}$ is the tree-level $\gamma$ and $Z$
boson exchange amplitude, which is the same in the CSI and the SM
and given by
\begin{eqnarray}
{\cal A}^{\gamma/Z}_{G^+G^-} =
-\frac{g_Z^2}{2}\left[\frac{s}{t}+\frac{t}{s}+1\right]\,.  
\end{eqnarray}
We have neglected the Higgs exchange diagrams since they are 
suppressed by $1/s$ or $1/t$. Thus we can explicitly see that the
tree-level amplitude in the SM,
\begin{eqnarray}
{\cal A}_{G^+G^-\rightarrow G^+G^-}^{\rm SM}=
-4\lambda_{\rm H}^{\rm SM}
-\frac{g_Z^2}{2}\left[\frac{s}{t}+\frac{t}{s}+1\right]\,,
\end{eqnarray}
agrees with the tree-level amplitude for $W^+W^-\rightarrow W^+W^-$ in
the high energy limit, {\it i.e.}, eq.\,\eqref{eq:WpWmSMlargeslim}
without the top-loop contribution.

Finally ${\cal A}^{s}_{G^+G^-}$ is the contribution of the
singlet-loop diagrams, which is given by
\begin{eqnarray}
  {\cal A}^{s}_{G^+G^-}=
  \frac{2N\lambda_{\rm HS}^2}{(4\pi)^2}
  \left[B_0(s;m_s^2)+B_0(t;m_s^2)\right]  \,,
\end{eqnarray}
with $s=(p_1+p_2)^2$ and $t=(p_1-k_1)^2$.  Using
eq.\,\eqref{eq:Dlambda_H}, it is straightforward to obtain
\begin{align}
  {\cal A}^{\rm CSI}_{G^+G^- \rightarrow G^+G^-}
  - {\cal A}_{G^+G^- \rightarrow G^+G^-}^{\rm SM} =
\frac{2N\lambda_{\rm HS}^2}{(4\pi)^2}\left[
B_0(s;m_s^2) + B_0(t;m_s^2)-2B_0(m_h^2;m_s^2)
\right]\,.
\label{eq:DAmpGpGm}
\end{align}

\subsection{$G^+G^+\rightarrow G^+G^+$ scattering}

\begin{figure}[t]
\begin{center}
 \includegraphics[scale=0.8]{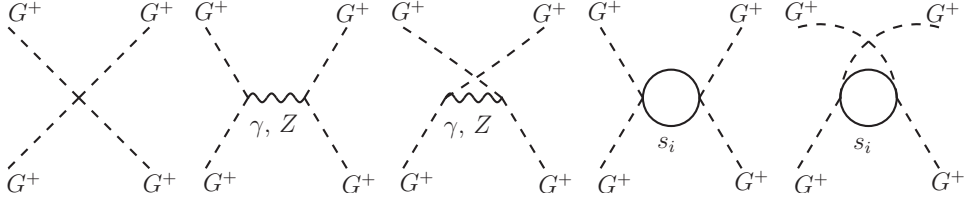}
\end{center}
\caption{\small Same as Fig.\,\ref{fig:AmpGpGm} but for
  $G^+G^+\rightarrow G^+G^+$ scattering. }
\label{fig:AmpGpGp}
\end{figure}

The same procedure can be used to derive the amplitude for
$G^+(p_1)G^+(p_2)\rightarrow G^+(k_1)G^+(k_2)$. The first three diagrams
in Fig.\,\ref{fig:AmpGpGp} give the SM amplitude,
\begin{eqnarray}
{\cal A}_{G^+G^+\rightarrow G^+G^+}^{\rm SM}=
-4\lambda_{\rm H}^{\rm SM}
-\frac{g_Z^2}{2}\left[\frac{u}{t}+\frac{t}{u}+1\right]\,,
\end{eqnarray}
which is exactly the same as the first two terms of
eq.\,\eqref{eq:WpWpSMlargeslim} as expected.  The last two diagrams
represent the additional contributions in the CSI model. They are
obtained similarly to the $G^+G^-\rightarrow G^+G^-$ case:
\begin{eqnarray}
  {\cal A}^{s}_{G^+G^+}=
  \frac{2N\lambda_{\rm HS}^2}{(4\pi)^2}
  \left[B_0(t;m_s^2)+B_0(u;m_s^2)\right]
\end{eqnarray}
with $u=(p_1-k_2)^2$, which leads to
\begin{align}
  {\cal A}^{\rm CSI}_{G^+G^+ \rightarrow G^+G^+}
  - {\cal A}_{G^+G^+ \rightarrow G^+G^+}^{\rm SM} =
\frac{2N\lambda_{\rm HS}^2}{(4\pi)^2}\left[
B_0(t;m_s^2) + B_0(u;m_s^2)-2B_0(m_h^2;m_s^2)
\right]\,.
\label{eq:DAmpGpGp}
\end{align}

\end{document}